\documentstyle[12pt,aaspp4,epsfig]{article}
\epsfverbosetrue

\def \ltaprx {\lower .1ex\hbox{\rlap{\raise .6ex\hbox{\hskip .3ex
	{\ifmmode{\scriptscriptstyle <}\else 
		{$\scriptscriptstyle <$}\fi}}}
	\kern -.4ex{\ifmmode{\scriptscriptstyle \sim}\else 
		{$\scriptscriptstyle\sim$}\fi}}}
\def\gtaprx {\lower .1ex\hbox{\rlap{\raise .6ex\hbox{\hskip .3ex
	{\ifmmode{\scriptscriptstyle >}\else 
		{$\scriptscriptstyle >$}\fi}}}
	\kern -.4ex{\ifmmode{\scriptscriptstyle \sim}\else 
		{$\scriptscriptstyle\sim$}\fi}}}

\def\sun{$_{\scriptscriptstyle_\odot}$}

\begin{document}

\title{COLLAPSARS - GAMMA-RAY BURSTS AND EXPLOSIONS IN ``FAILED SUPERNOVAE''}

\author{A. MacFadyen and S. E. Woosley}
\affil{Astronomy Department, University of California, Santa
  Cruz, CA 95064}
\affil{e-mail: andrew@ucolick.org; woosley@ucolick.org}

\lefthead{MACFADYEN AND WOOSLEY}
\righthead{COLLAPSARS}

\date{\today}

\begin{abstract}

Using a two-dimensional hydrodynamics code (PROMETHEUS), we explore
the continued evolution of rotating helium stars, M$_{\alpha} \gtaprx
10$ M\sun, whose iron core collapse does not produce a successful
outgoing shock, but instead forms a black hole of 2 - 3 M\sun. The
model explored in greatest detail is the 14 M\sun \ helium core of a
35 M\sun \ main sequence star. The outcome is sensitive to the angular
momentum. For $j_{16} \equiv j/(10^{16}$ cm$^2$ s$^{-1}) \ltaprx$ 3,
material falls into the black hole almost uninhibited. No outflows are
expected. For $j_{16} \gtaprx 20$, the infalling matter is halted by
centrifugal force outside 1000 km where neutrino losses are
negligible. The equatorial accretion rate is very low and explosive
oxygen burning may power a weak equatorial explosion. For 3 $\lesssim
j_{16} \lesssim $ 20, however, a reasonable value for such stars, a
compact disk forms at a radius where the gravitational binding energy
can be efficiently radiated as neutrinos. These are the best
candidates for producing gamma-ray bursts (GRBs). Here we study the
formation of such a disk, the associated flow patterns, and the
accretion rate for disk viscosity parameter, $\alpha
\approx 0.001$ and 0.1. Infall along the rotational axis is initially
uninhibited and an evacuated channel opens during the first few
seconds.  Meanwhile the black hole is spun up by the accretion (to $a
\approx 0.9$) and energy is dissipated in the disk by
magneto-hydrodynamical (MHD) processes and radiated by neutrinos. For
the $\alpha = 0.1$ model, appreciable energetic outflows develop in
cones with polar angle about 30 - 45 degrees. These outflows, powered
by dissipation in the disk, have energy up to a few times 10$^{51}$
erg, mass $\sim$ 1\,M\sun, and are rich in $^{56}$Ni. They constitute
a supernova-like explosion by themselves. Meanwhile accretion through
the disk is maintained for at least 20 s, but is time variable
($\pm$30\%) because of hydrodynamical instabilities at the outer edge
in a region where nuclei are experiencing photodisintegration. Because
the efficiency of neutrino energy deposition is sensitive to the
accretion rate, this instability leads to highly variable energy
deposition in the polar regions. Some of this variability, which has
significant power at 50 ms and overtones, may persist in the time
structure of the burst. During the time followed, the average
accretion rate for the standard $\alpha$=0.1 and $j_{16}$ = 10 model
is 0.07 M\sun \ s$^{-1}$ and the total energy deposited along the
rotational axes by neutrino annihilation, (1 - 14) $\times 10^{51}$
erg, depending upon the evolution of the Kerr parameter and uncertain
neutrino efficiencies. Simulated deposition of this energy in the
polar regions, at a constant rate of $5 \times 10^{50}$ erg s$^{-1}$
per pole, results in strong relativistic outflow - jets beamed to
about 1.5\% of the sky. The jets blow aside the accreting material,
remain highly focused, and are capable of penetrating the star in 5 -
10 s. After the jet breaks through the surface of the star, highly
relativistic flow can commence.  Because of the sensitivity of the
mass ejection and jets to accretion rate, angular momentum, and disk
viscosity, and the variation of observational consequences with
viewing angle, a large range of outcomes is possible ranging from
bright GRBs like GRB 971214 to faint GRB-supernovae like SN
1998bw. X-ray precursors are also possible as the jet first breaks out
of the star. While only a small fraction of supernovae make GRBs, we
predict that all GRBs longer than a few seconds will make supernovae
similar to SN 1998bw.  However, hard, energetic GRBs shorter than a
few seconds will be difficult to make in this model.

\end{abstract}

\vfill\eject

\section{INTRODUCTION}

Despite 60 years of speculation (e.g., Baade \& Zwicky 1934; Hoyle
1946) and 30 years of intensive calculation (e.g., Fowler \& Hoyle
1964; Colgate \& White 1966; Arnett 1967; Wilson 1971) the exact
mechanism whereby the collapsing iron core of a massive star produces
an outgoing shock and makes a supernova remains uncertain. Controversy
has surrounded this subject since the first computer models were
published in the late 1960's (Colgate 1968; Arnett 1968). Modern
calculations (Herant et al. 1994; Burrows, Hayes, \& Fryxell 1995;
Janka \& M\"uller 1996; Fryer, Benz, \& Herant 1996) suggest that the
explosion is powered by neutrino energy deposition in a hot,
convectively unstable bubble of radiation and pairs just outside the
proto-neutron star. Most of these calculations show an explosion
developing in two-dimensional models using approximate neutrino
physics. But their success is challenged (e.g., Mezzacappa et
al. 1998) and even should this mechanism work for some stars, it may
fail for others (Burrows 1998; Fryer 1998), especially the more
massive ones. This is because more massive stars have denser, thicker
mantles of oxygen and silicon overlying the collapsing iron
core. These mantles provide high accretion rates and ram pressure that
are difficult for the hot bubble to overcome. For some mass of star,
often speculated to be around 25 to 35 M\sun \ on the main sequence
(helium core 9 to 14 M\sun), the protoneutron star accretes enough
matter before an explosion develops that it becomes a black
hole. After that, the compact object no longer radiates
neutrinos. Such calculations are often termed ``failures'' by those
who carry them out because they don't get a supernova, at least not
the way they expected. In this paper, lacking definite calculations,
we postulate the existence of such failures and explore their
continued evolution. It turns out they are not such failures after
all.

Without rotation, this evolution is simple. The star falls into the
black hole in a hydrodynamical time scale, carrying any internal
energy with it, and simply disappears. In nature, it is doubtful that
this ever occurs. The outer layers of the star, and, in at least some
cases, the mantle have too much angular momentum to fall freely inside
the last stable orbit. An accretion disk forms where the dissipation
of rotational and gravitational energy will give rise to some sort of
mass ejection and electromagnetic display, though, as we shall see, a
lot of the energy may come out as neutrinos.

The study of ``Failed supernovae'' was initiated by Bodenheimer \&
Woosley (1983) and the model has been explored, in a preliminary way,
as a GRB progenitor by Woosley (1993; 1996), Hartmann \& Woosley
(1995), Jaroszy\'nski (1996), and Popham, Woosley, \& Fryer (1998;
henceforth PWF). Paczy\'nski (1998) has discussed some of the
observational consequences of the collapsar model in a phenomenon he
calls the ``hypernova''.

In this paper we study the evolution of such objects in much greater
detail than previous works using a two-dimensional hydrodynamics code
and more realistic inner boundary conditions and disk physics than,
for example, Bodenheimer \& Woosley. We survey the effect of different
values of angular momentum and disk viscosity and also explore the
consequences of energy transport from the accreting disk by neutrinos
or postulated MHD effects. We start with a collapsing star removing
the assumption of ``stationarity'' (e.g. Jaroszy\'nski) and follow the
formation of the accretion disk and it's subsequent evolution.  While
our paper will focus on the evolution of bare helium stars whose
iron cores collapse to black holes, there are other ways of reaching
similar initial conditions, especially the merger of a black hole with
the helium core of red supergiant star following common envelope
evolution (Fryer \& Woosley 1998; PWF) and white dwarf-black hole
mergers (Fryer et al. 1998).  Our model, though motivated by the
desire to make a GRB, has the potential to create a strong
supernova-like outburst or both. As we shall discuss ($\S$6.2), SN 1998bw
(Galama et al. 1998) may have been an example.

\section{THE INITIAL MODEL}

Besides the prompt formation of a black hole, the other essential
ingredient in our model is rotation. Specific angular momentum,
$j_{16}$, of at least a few is needed so that a disk will form well
outside the last stable orbit for a black hole of several solar
masses.  For a Schwarzschild black hole the radius of the last stable
circular orbit is $r_{\rm lso} = 2.7 \times 10^6 (M_{bh}/3M_{\odot})$
cm and this orbit has specific angular momentum $ j_{\rm lso,16} = 4.6
(M_{bh}/3M_{\odot})$.  The corresponding values for a rotating black
hole with Kerr parameter $a = 0.95$ are $r_{\rm lso} = 8.6 \times 10^5
(M_{\rm bh}/3M_{\odot})$ cm and $ j_{\rm lso,16} = 2.5
(M_{\rm bh}/3M_{\odot})$; for $a=1$, $j_{\rm lso,16} = 1.5 (M_{\rm
bh}/3M_{\odot})$. General expressions are given in $\S$4.1.6.

\begin{figure}
\vspace{1 in}
\epsfxsize=15 true cm
%\epsffile[0 0 612 612]{heger.ps}
\epsffile[0 0 612 612]{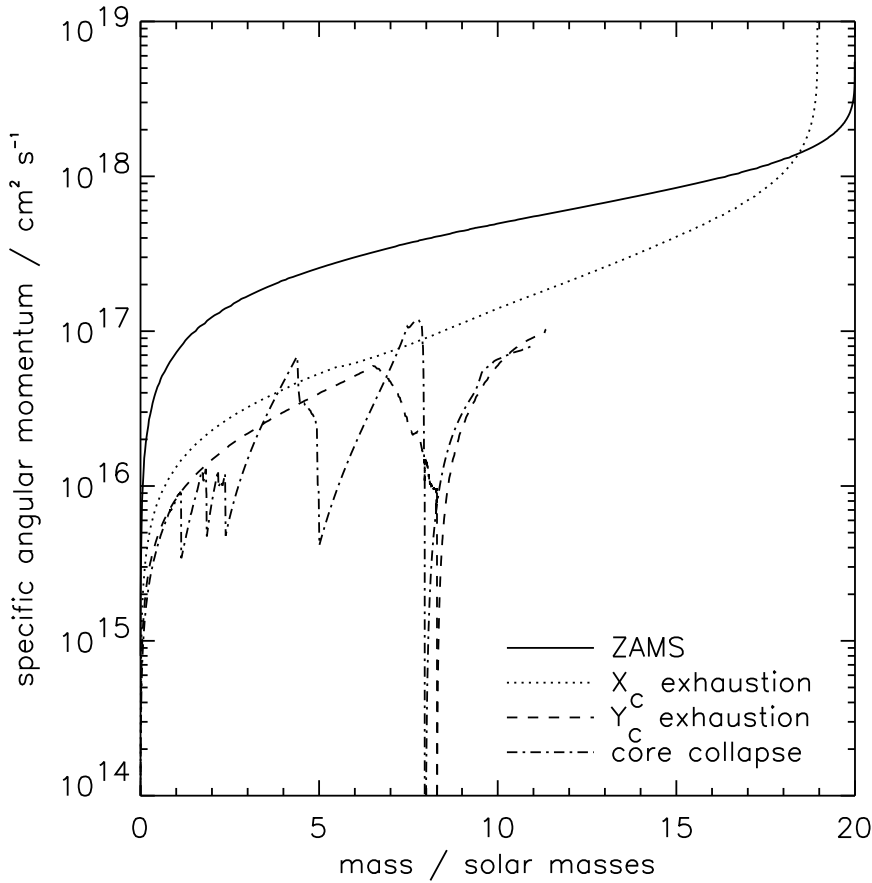}
\vspace{-1.5 in}
\caption{\label{heger} Angular momentum ($j$) distribution in a 20
M\sun \ star evolved from an initially rigidly rotating main sequence
star with equatorial velocity 200 km s$^{-1}$ by Heger, Langer, \&
Woosley (1998). The solid line gives $j$ on the main sequence; the
dotted line is at hydrogen depletion; the dashed line, at helium depletion;
and the dash-dotted line is the presupernova star. A decrease of the
total stellar mass by mass loss during hydrogen and helium burning is
apparent. The distribution of $j$ in the presupernova star shows sharp
decreases at the outer edges of convective shells and the helium core
mass is 8 M\sun.  The helium cores considered here are more massive, but may
have a similar distribution of angular momentum. The numbers plotted
here have been averaged over spherical shells and the actual
equatorial angular momentum is about 50\% higher.
}

\end{figure}

Angular momenta of this magnitude, and more, are characteristic of
current presupernova models in the mass range 10 - 20 M\sun and may
also characterize more massive stars. Fig. \ref{heger} shows the
calculated distribution of $j$ for a 20 M\sun \ star evolved by Heger,
Langer, \& Woosley (1998). The central angular momentum is about an
order of magnitude less than what one would exist had angular momentum
had been conserved in the core all the way from a (rigidly rotating)
main sequence model with typical observed rotation speed (about 200 km
s$^{-1}$). However, the calculated presupernova angular momentum in
the stellar core is still about two orders of magnitude greater than
observed even in fast pulsars like the 16 ms pulsar in SNR N157B
(Marshal et al. 1998).  Perhaps pulsars are slowed during or after the
supernova explosion (Lindblom et al. 1998; Owen et al. 1998). But
magnetic fields have been ignored in the Heger et al.  calculations.
If the helium core is braked by a magnetic field prior to the
supernova explosion to the extent described by Spruit \& Phinney
(1998), then our model will not work for single stars. One would need
to invoke the late time merger of a close binary (e.g., Paczy\'nski
1998) or the black hole helium-core mergers discussed by Fryer \&
Woosley (1998). Because we are considering an event that happens
at $\ltaprx$1\% of the Type II supernova rate, such rare occurrences
would be acceptable.

The formation of the massive, rapidly rotating helium
stars desired here is probably favored by low metallicity. Low metallicity
keeps the radius of the star smaller and also reduces the mass
loss. Both effects inhibit the loss of angular momentum by the star.
One might then need a close binary to remove the envelope and make the
assumed bare helium core (see below), but that condition is not very
restrictive. The mechanism whereby helium cores (Wolf-Rayet stars)
continue to lose mass after their envelopes are gone is uncertain
(e.g., Langer 1989), but that too might be sensitive to metallicity. By
raising the threshold for removal of the hydrogen envelope by stellar
winds, low metallicity also increases the mass of the heaviest
helium core one makes in a given generation of stars and thus favors
black hole production.

{\sl This dependence on metallicity implies a possible evolution of GRB
characteristics with red shift.}

As we shall see, though perhaps less obviously, it is also important
that the GRB progenitor not have {\sl too much} angular momentum. For
angular momenta $j_{16} \gtaprx 15$, the accretion disk forms far
outside of several hundred km in a region where neutrino losses are
unimportant. Lacking this efficient means of energy dissipation, it is
difficult to form a bound disk. The resulting flow patterns are
different and favor outflow ($\S$4.3). Most importantly, the accretion
rate into the hole is reduced. High accretion rate is essential if the
burst is to be powered by neutrinos (PWF).

All stars are assumed to have lost their hydrogen envelopes. This may
require binary membership for stars of solar metallicity and main
sequence mass under about 30 M\sun \ (though see Heger, Langer, \&
Woosley 1998), but for lower metallicity, the mass limit is
higher. Should the star retain its hydrogen envelope, an explosion of
the sort we shall describe would still develop with interesting
observational consequences but, during the time the black hole
accreted at a rate high enough to make a GRB, the hydrogen envelope
would remain stationary. It would be difficult for a jet with a
significant opening angle to retain a high relativistic $\Gamma$ while
plowing though the overlying matter. Our helium cores have a radius of
less than a light second and, as we shall see, a sustained relativistic jet
can punch a hole through the star. This would not be the case for a star with
radius 1000 light s, that is, a red supergiant. However, compact WN stars
would serve our purpose just as well. A layer of surface hydrogen is allowed
so long as its radius is not large. A compact star is also favored by low
metallicity.

For our calculations, we use the helium cores of 25 M\sun
\ and 35 M\sun \ presupernova stars (Woosley \& Weaver 1995). These
stars were evolved from the main sequence, without mass loss or
rotation, to the presupernova star using the KEPLER stellar evolution
code (Weaver, Zimmerman, \& Woosley 1978). Since we will be interested
chiefly in the evolution of the deep interior of these stars, the
treatment of the surface is not so important. We extracted the helium
cores of these stars as defined by the point where the hydrogen mass
fraction declined below 0.01. We call this the ``helium core mass'',
M$_{\alpha}$. Various calculations used different fractions of the
helium core mass, but usually the whole core was carried. 

\begin{figure}
\epsfig{file=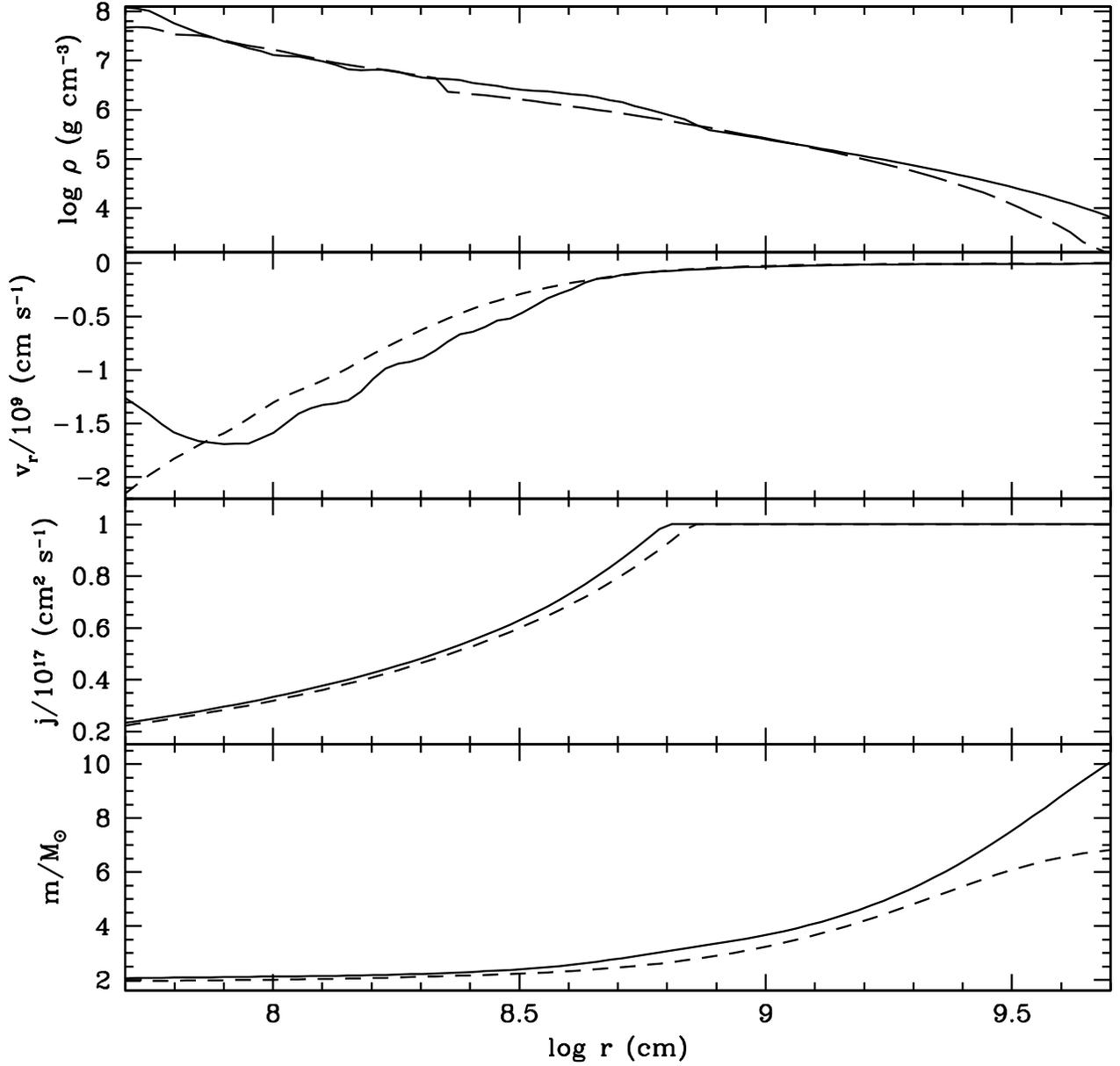,width=7 in}
\caption{\label{initmod} Density, radial velocity, specific angular
momentum, and enclosed mass in the equatorial plane of 
the two initial models used for this
study. The solid line shows Model 14A derived from a 35 M\sun \
presupernova star (Woosley \& Weaver, 1995). The dashed line shows
Model 9A derived from a 25 M\sun \ presupernova model. The former is
our standard case. The material with low $j$ falls in very early and
most of the calculation was for $j \approx$ constant $\approx$
10$^{17}$ cm$^2$ s$^{-1}$. }
\end{figure}

For the core derived from the 25 M\sun \ presupernova, M$_{\alpha}$ =
9.15 M\sun \ and the iron core was 1.78 M\sun. At the time Woosley \&
Weaver (1995) defined as the ``presupernova'' (collapse velocity equal
1000 km s$^{-1}$), the radius of this core was still 2300 km.
This inner boundary was moved smoothly into either 50 or 200 km before
beginning our calculation. At that point, the collapse velocity,
density, and assumed specific angular momentum are given in
Fig. \ref{initmod}. See Woosley \& Weaver (1995) for details of the
composition which is mostly oxygen and helium. 

A similar model of M$_{\alpha}$ = 14.13 M\sun \ was generated from a
35 M\sun \ presupernova model. The collapse velocity and angular
momentum distributions were similar to the 9 M\sun \ model, but the
density declined more slowly with radius (Fig. \ref{initmod}). The
mass of the iron core removed was 2.03 M\sun. This became our standard
Model 14A. Another model was explored with lower disk viscosity (Model
14B; $\S$4.2).

Angular momentum was distributed so as to provide a constant ratio of
.04 of centrifugal force to the component of gravitational force
perpendicular to the rotation axis at all angles and radii, except
where that prescription resulted in $j_{16}$ greater than a prescribed
maximum.  In most cases (all but $\S$4.4) the maximum value of
$j_{16}$ was 10.  Thus in all cases the ratio of centrifugal support
to gravity was small and the use of a presupernova model that had been
calculated without rotation was justified. This maximum value of
$j_{16}$ is consistent with the presupernova calculations shown in
Fig. \ref{heger}, though larger by about 50\%. Since the inner 2 to
2.5 M\sun \ of the star collapsed very rapidly into the inner boundary
(i.e., the assumed black hole), the exact value of angular momentum
there did not matter much, except as it influenced the initial Kerr
parameter of the black hole.  Future studies will explore the
sensitivity of our results to the assumed distribution of angular
momentum.

Most of our studies, including Models 14A and 14B, used a perfectly
absorbing inner boundary condition at 50 km. The smaller the radius of
the inner boundary, the more restrictive is the Courant condition on
the time step and the more computer time one must spend to evolve to a
given epoch. This radius was a reasonable compromise between what
could be computed and the fact that most of the interesting physics
went on inside of several hundred km. At the inner boundary pressure
turned out to be about 15\% of gravity, consistent with analytic
models by PWF. Centrifugal forces dominate the force balance of the
inner disk, which turns out to be thin because of efficient neutrino
cooling. Thus the use of an absorbing inner boundary is justified.

\section {THE PPM CODE AND ITS MODIFICATIONS}

The presupernova models, which were already collapsing at a few
thousand km s$^{-1}$ at the time the link was made, were mapped onto an
Eulerian grid and the subsequent evolution followed using PROMETHEUS
(Fryxell, M\"uller, \& Arnett 1989, 1991; M\"uller 1998), a two-dimensional
hydrodynamics code based upon the Piecewise Parabolic MUSCL scheme (PPM;
Woodward \& Collela 1984).  Axial symmetry and reflection symmetry across the
equatorial plane were assumed. Spherical coordinates $(r,\theta)$ were
employed with logarithmic zoning in the radial direction and regular zoning
in $\theta$. Typically 150 radial zones and 27 angular zones were used. The
total number of zones was thus $\sim$4000. This relatively sparse grid was
necessary because of the large number of time steps imposed by the Courant
condition at small radii.

The PROMETHEUS code was modified to include a realistic equation of
state (EOS; Blinnikov, Dunina-Barkovskaya, \& Nadyozhin, 1996) which
included a Fermi gas of electrons and positrons - with arbitrary
relativity and degeneracy, radiation, and a Boltzmann gas of nuclei.
The necessary Fermi integrals were done using analytic expressions that
allowed a quick solution without the use of extensive tables.  Coulomb
corrections were included for densities above 10$^4$ g cm$^{-3}$
(Shapiro \& Teukolsky 1983). The KEPLER composition was mapped into
nine species: neutrons, protons, helium, carbon, oxygen, neon,
magnesium, silicon, and nickel. Except in regions where
photodisintegration was important (see below), the original
composition of KEPLER was preserved and simply advected by the
hydrodynamics code.  Everywhere, even in the nucleonic disk, a
constant electron mole number, $Y_e$ = 0.50 was assumed. In the inner
disk electron capture may decrease $Y_e$. While interesting for
nucleosynthesis, this detail was not important for calculating the
thermodynamic properties of the disk.

Nuclear processes such as carbon, neon, oxygen, and silicon burning,
and electron-capture were not followed in this first study except for
one model where oxygen burning was implemented using an analytic
formula ($\S$4.3). The code we constructed included a 9-isotope
nuclear reaction network capable of following all these burning
processes, but because of the restrictive time steps its operation
imposed, it was turned off.  However, the dominant nuclear energy term
here is the photodisintegration of helium and heavier elements into
neutrons and protons.  A simplified treatment of captured the
essential effects.  Photodisintegration was incorporated directly into
the EOS by including the the nuclear binding energy (with zero point
set at pure $^{56}$Ni) as part of the energy density.  Nuclear
statistical equilibrium (NSE) was assumed to compute the free nucleon
mass fraction at a given temperature and density (Woosley \& Baron
1992): $$\rm X_{nucleon} = 26 T_{MeV}^{9/8} / \rho_{10}^{3/4}
exp(-7.074/T_{MeV}). $$

This makes the assumption (valid at high temperature) that the time
scale to reach and maintain equilibrium is much shorter than the
hydrodynamical time.  Each time the EOS is called with a new total
energy density (thermal plus nuclear binding) and mass density, a
Newton-Rapheson iteration is performed over temperature to
simultaneously solve for the new thermodynamic variables (temperature,
pressure, entropy, $\Gamma_1$) and the new free nucleon mass fraction.
The nuclear physics was further simplified by treating only the
transition from ``heavy nuclei'' ($\rm A > 1$) to free nucleons.
Transitions among the heavy nuclei and from heavy nuclei to
alpha-particles were neglected since roughly 90\% of the energy loss
to photodisintegration occurs when helium disintegrates to free
nucleons.  The ``heavy nuclei'' abundances were renormalized to make
the sum of their mass fractions equal to 1 - X$_{nucleon}$ each time
X$_{nucleon}$ was calculated.

Effects of viscosity in the disk were implemented using the alpha
viscosity prescription of Shakura and Sunyaev (1973), $ \nu = \alpha
c_{s} H$, where $c_{s}$ is the local sound speed and $H$, the density
scale height.  The full stress tensor was calculated (Tassoul 1978)
and appropriate terms included in the momentum and energy equations.
For simplicity, all terms in the viscous stress tensor except the
$\tau_{\phi r}$ terms were set to zero for these calculations.  The
density scale height was calculated along arcs of constant radius by
determining the angular zone where the density first dropped a factor
of $e$ below the equatorial value.  Since the disk studied here is
embedded in a collapsing stellar envelope, it was desirable to
implement disk viscosity only in the regions of the simulation where
the flow had become disk-like.  Viscosity was turned on smoothly for
zones in regions of approximate radial force balance, i.e., those
which were making at least a few orbits before accreting.
Specifically, $\alpha$ was modified as follows, if abs $(v_r/v_{\phi})
> 1$, $\alpha = \alpha_0$ min $(1,(0.1 v_r/v_{\phi})^2)$, and if
abs\,$(v_r/v_{\phi}) < 1$, $\alpha = 0$.

Since viscous dissipation can become very large in the inner disk
($\gtaprx 10^{30}$ erg cm$^{-3}$ s$^{-1}$), viscous heating and
neutrino cooling terms were included together in the energy
equation. Subcycling was implemented that allowed the hydrodynamical
time step to be used wherever possible.  Additional constraints were
set on the time step to limit the total changes in energy, temperature
and abundances to less than a few percent per step.  In practice
however, the Courant condition was used for a majority of the time
steps.

Neutrino losses in the optically thin limit were included with thermal losses
(dominated by pair annihilation) taken from Itoh et. al. (1989, 1990) Neutrino
emission due to pair capture on free nucleons were also included
using an approximation
$$\epsilon_{\rm pair-cap} = 9 \times 10^{33} \, T_{11}^6 \, \rho_{10}
X_{\rm nucleon} \ \ {\rm erg \ cm^{-3} \ s^{-1}},$$ where $ T_{11} =
T/10^{11}$ K, $\rho_{10} = \rho/10^{10}$ g cm$^{-3}$, and
X$_{nucleon}$ is the free nucleon mass fraction given above.
Neutrinos from pair capture are an important energy sink in
the hot, dense parts of the torus where the nuclei have disintegrated
into free nucleons and are generally more important than neutrinos
from pair annihilation.

Poisson's equation for the gravitational potential was solved using an
integral solver (M\"uller \& Steinmetz 1995).  For two dimensions in
spherical coordinates this solver is computationally efficient using
only $\sim1\%$ of the computation time.  The gravitational potential
of the central point mass was modified to account for some of the
effects of general relativity (Paczy\'nski \& Witta 1980): $$\phi \ =
\ - GM/(r-r_s),$$ where $r_s = 2GM/c^2 $ This potential reproduces the
positions of the last stable circular orbit and marginally stable
circular orbit and approximately reproduces the binding energy of the
last stable orbit.  In our calculations the inner boundary, 50 km, was
always greater than $\sim 4$ Schwarzschild radii, so this should be
sufficiently accurate.  The point mass was increased during the
calculation by the amount of baryonic mass which flowed across the
inner boundary.  The gravitational mass of the hole is probably
smaller than the baryonic value by $\sim$ 5\% due to neutrino emission
from the inner disk.  This effect was not included in computing the
potential, but shouldn't be a large effect.

\section{COLLAPSE AND DISK DYNAMICS}

\subsection {The standard model}

The evolution of Model 14A, as previously defined, was followed for an
elapsed time of 20 seconds (nearly 2 million time steps). Its evolution
can be considered in three stages. 

First is a transient stage lasting roughly 2 seconds, during
which low angular momentum material in the equator and most of the
material within a free fall time along the axes falls through the
inner boundary. A centrifugally supported disk forms interior to
roughly 200 $(j_{16}/10)^2$ km. The density near the hole and along
its rotational axis drops by an order of magnitude.

The second stage is characterized by a quasi-steady state in which
the accretion disk delivers matter to the hole at approximately the same rate
at which it is fed at its outer edge by the collapsing star. The
average accretion rate, about 0.07 M\sun \ s$^{-1}$, is slower than expected
simply from free fall, $M/\tau_{\rm HD} \approx M \bar \rho^{1/2}/446 \approx
1$ M\sun \ s$^{-1}$ for $\bar \rho$ = 10$^4$ g cm$^{-3}$ and M = 10 M\sun,
because pressure remains important in the star even though its core has
collapsed.  This stage of enduring rapid accretion at an approximately
constant rate is the most interesting one for making a GRB. Large energy
deposition can occur in the polar regions by neutrino annihilation and MHD
processes. However, as we shall see, the GRB cannot commence until the mass
density in the polar regions falls below a critical value, about 10$^6$ g
cm$^{-3}$. The GRB producing stage, if it is going to happen, thus starts
several seconds after the initial collapse and continues for at least another
15 s after which the accretion rate begins to decline. If the energy
deposition by neutrinos and MHD processes occurs at too slow a rate, jet
formation may be delayed until most of the accretion and energy generation is
over.

The third stage is the explosion of the star. This occurs on a longer
time scale and we were not able to follow it all the way. Energy deposited
near the black hole along the rotation axes makes jets that blow aside what
remains of the star within about 10 - 20 degrees of the poles, typically
$\sim$0.1 M\sun. The kinetic energy of this material pushed aside is quite
high, a few $10^{51}$ erg, enough to blow up the star in an axially driven
supernova. Additional energy is deposited by viscous processes, presumably
MHD in nature, in and above the disk. This also gives high ejection
velocities to larger amounts of mass at larger angles ($\S$4.1.5). During the
tens of seconds that it takes the star to come apart, if energy input
continues at their base, the relativistic jets created in the deep interior
erupt from the surface of the star and break free. Their relativistic
$\Gamma$ rises. They then travel hundreds of AU's before making the GRB.

We now consider each stage in greater detail.

\subsubsection {Disc formation}

All gas with angular momentum less than the Keplerian value at the 50
km inner boundary, $j_{16} < 4.6 \, \sin^2(\theta) \, ({\rm M_{bh}}/3
\ {\rm M_{\scriptscriptstyle \odot}})^{1/2}$, can fall uninhibited
through the inner boundary at polar angle $\theta$, though for angular
momenta larger than the last stable orbit ($\S$4.1.6), a disk may
still form interior to that boundary.  As soon as gas with larger $j$
reaches the inner boundary, a centrifugally supported torus starts to
form with a surrounding accretion shock. Fig. \ref{accshock.j} shows
this accretion shock at a time, 0.751 s, when it is moving out rapidly
in both mass and radius.  Later it becomes more spherical. Here
centrifugal force balances gravity at about 200 km. The temperature
and density interior to the accretion shock are $\gtaprx 10^{10}$ K
and $\gtaprx 10^8$ g cm$^{-3}$.  At these temperatures the accreting
gas, mostly oxygen and silicon, photodisintegrates into free neutrons
and protons (Fig. \ref{accshock.abar}).  The neutrino emission, which
outside the photodisintegration region is dominated by pair
annihilation, is greatly enhanced in this inner region by the capture
of abundant electron-positron pairs onto neutrons and protons.  For
all densities encountered {\sl on our grid} in Model 14A (but not
Model 14B) the gas is optically thin to neutrinos and the neutrino
emission was treated as a local energy loss (though see $\S$4.1.7).
 
After two seconds, phase 1 is ending. The accretion shock has moved
out to 8,500 km and is roughly spherical.  The temperature behind this
shock has declined to $\approx 2.5 \times 10^9$ K and
photodisintegration there has ceased.  However, a second accretion
shock now bounds the disk at 250 km. Partial photodisintegration
occurs at about 1000 km due to adiabatic compression, but most of the
energy is absorbed as full photodisintegration occurs in the disk
shock (though see $\S$4.1.4).  The polar accretion velocity is
approaching $c/2$ (the free fall speed) at 60 km, but the equatorial
density is already three orders of magnitude higher ($8 \times 10^8$ g
cm$^{-3}$) and accretion through the equator dominates the total
accretion rate even though the equatorial accretion velocities (20,000
km s$^{-1}$) are seven times smaller.

\begin{figure} 
\epsfig{file=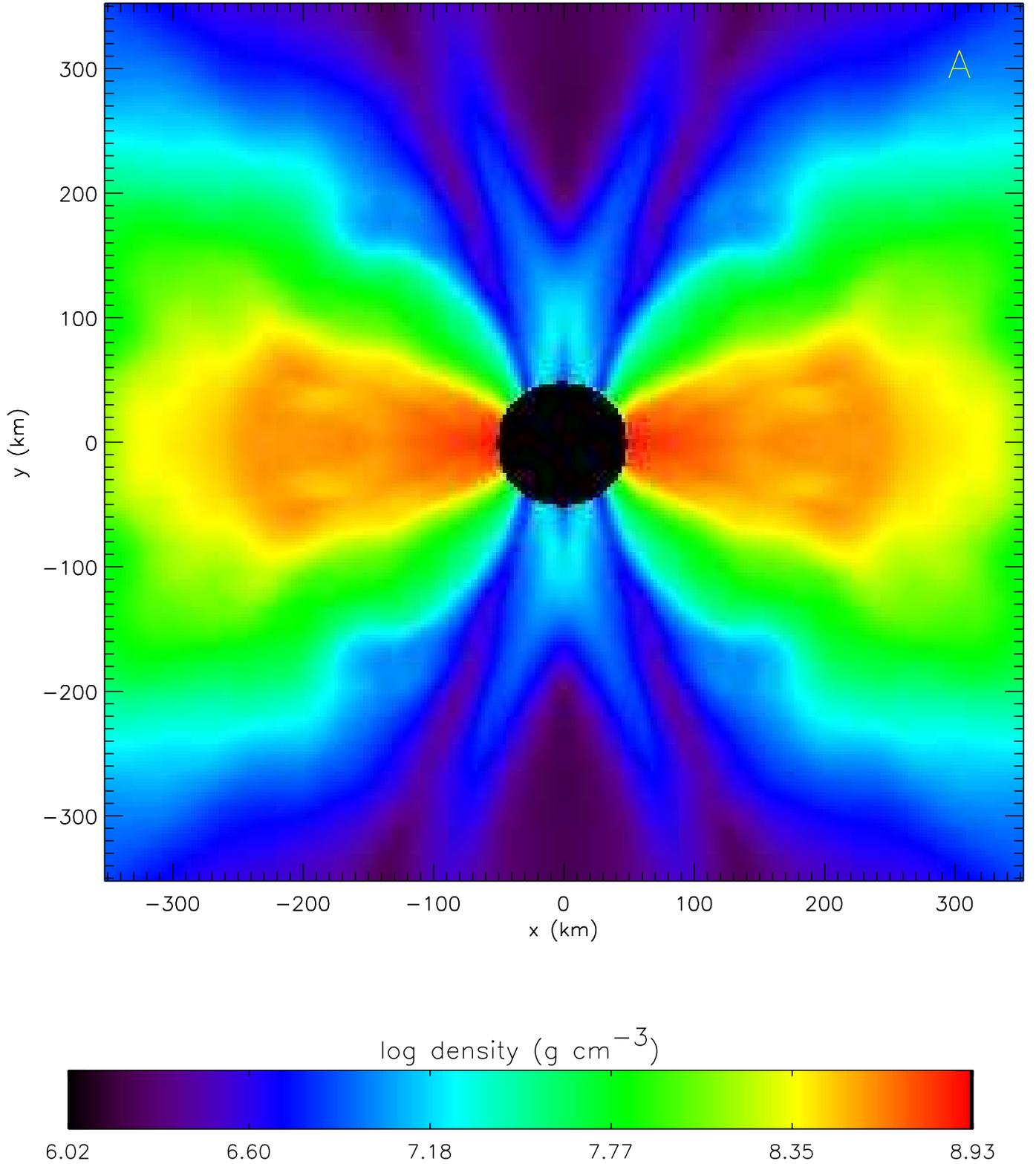}
\caption{\label{dens14a.small} The density in the central regions of
Model 14A 7.60 seconds after core collapse.  A dense disk (red, 10$^9$
g cm$^{-3}$) of gas is accreting into the black hole. The
centrifugally supported torus has a radius of 200 km. Still higher
densities exist in the disk inside the inner boundary of our
calculation (50 km). Gas is accreting much more readily along the
polar axis due to the lack of centrifugal support and has left behind
a channel with relatively low density (blue, 10$^6$ g cm$^{-3}$). Should
energy be deposited near the black hole, this geometry will
naturally focus jets along the rotational axis.}
\end{figure}

This large density contrast which has already begun to develop between
the poles and the equator is very important for the viability of the
collapsar as a GRB model.  Subsequent evolution increases this
contrast (Fig. \ref{dens14a.small}).  This hourglass geometry is quite
favorable for the geometrical focusing of jets.

\subsubsection {The steady state disk}

\begin{figure} 
\epsfig{file=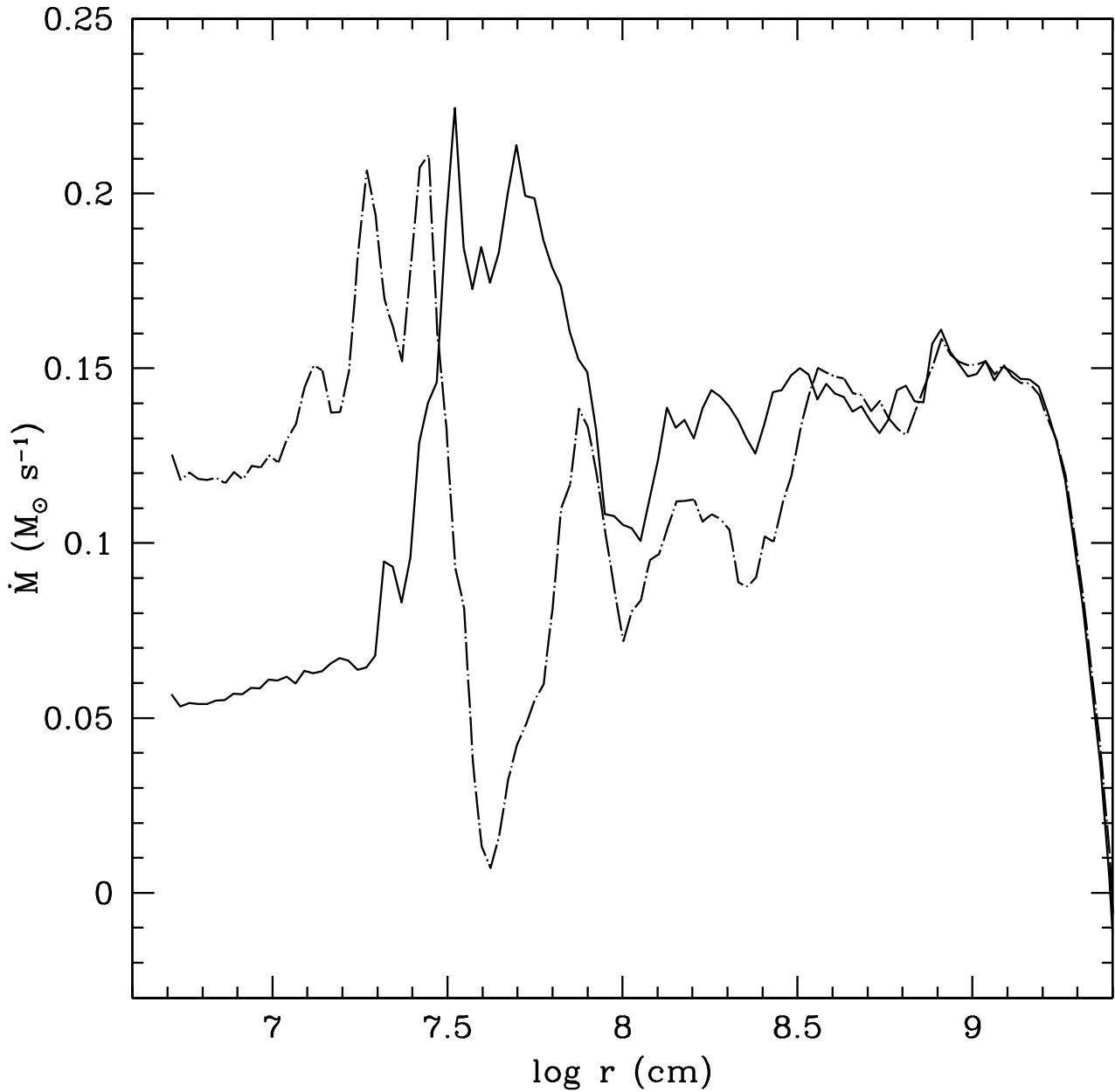,width=7 in}
\caption{\label{mdotofr} The mass accretion rate as a function of
radius within 45 degrees of the equator. Farther out, where most of
the mass is, the rate is just given by the dynamical response of a
star that has had its central pressure support removed. The accretion
rate is about 0.14 M\sun \ s$^{-1}$, but roughly half as much is
flowing outwards at polar angles between 15 and 40 degrees
(Fig. \ref{plumes}). Closer to the disk centrifugal forces, shocks,
photodisintegration, and multidimensional flows come into play and the
accretion rate varies from about 0.06 M\sun \ s$^{-1}$ (solid line;
``low" state; t = 7.540 s) to 0.12 M\sun \ s$^{-1}$ (dash-dot line;
``high" state; t = 7.598 s). }
\end{figure}

After a few seconds, a quasi-steady state exists for the accretion
disk. Matter supplied through an accretion shock at about 200 - 300 km
is transported by viscous interaction to the inner boundary at about
the same rate at which it passes through the shock
(Fig. \ref{mdotofr}). Interesting deviations from this steady state
exist outside the inner accretion shock, but the disk responds promptly
to these variations and between 50 and 200 km, mass flux is very
nearly constant.
The steady state disk for Model 14A has a low mass, a few thousandths
of a solar mass. Later we shall see that the mass of the disk varies
roughly inversely with the viscosity parameter, $\alpha$, and can
become much larger for inviscid disks (PWF and $\S$4.2).

Figure \ref{popham} shows the physical conditions in the equatorial
plane of Model 14A at a time 7.598 s after core collapse when the
accretion rate is 0.12 M\sun \ s$^{-1}$ and the black hole mass 3.5
M\sun. The density, temperature, rotation rate, radial velocity and
angular momentum are all shown as a function of radius for the inner
10,000 km of the problem (the outer boundary of the grid was at 50,000
km). Also shown is the density scale height, H, where the
equatorial density declines by a factor of $e$. All of these
quantities are compared with
the semi-analytic solution of PWF. The latter is a steady-state
one-dimensional "slim disk" solution for a 3 solar mass Schwarzschild
black hole (a = 0), with viscosity parameter, $\alpha = 0.1$, accreting at
0.1 M\sun \ s$^{-1}$.  The PWF model also included terms in the EOS
to represent approximately the effects of electrons (degenerate and
non-degenerate) and pairs though our EOS (Blinnikov et al. 1996) is
more accurate and general.  Photodisintegration and neutrino emission
were treated in a similar way in both studies. However, the PWF
calculation was one dimensional (the disk was vertically averaged) and
assumed steady state. Its great strength was its ability to follow
disks, for various choices of accretion rate, disk viscosity, and hole
mass, into the deepest regions where most of the energy is released
and general relativity is increasingly important, especially for
rapidly rotating black holes.

The good agreement with PWF, in the region where a steady state disk
ought to exist (interior to 200 km), serves to mutually validate both
calculations and to verify the steady state assumption. The accretion
shock is apparent in the radial velocity plot of our new
results. Outside that shock, one expects and sees major differences
with PWF. For example, the density plot also shows an inversion, a
torus, at 200 km where infalling material piles up in the
multi-dimensional study. However, the radial velocities in the disk
agree very well verifying that both calculations implemented $\alpha =
0.1$ consistently.  The temperature is especially well replicated as
is the thinning of the accretion disk, both inside and outside the
accretion shock, as a consequence of photodisintegration. The disk
interior to a few hundred km has a scale height about 40\% of the
radius and is ``slim''.

\begin{figure} 
\epsfig{file=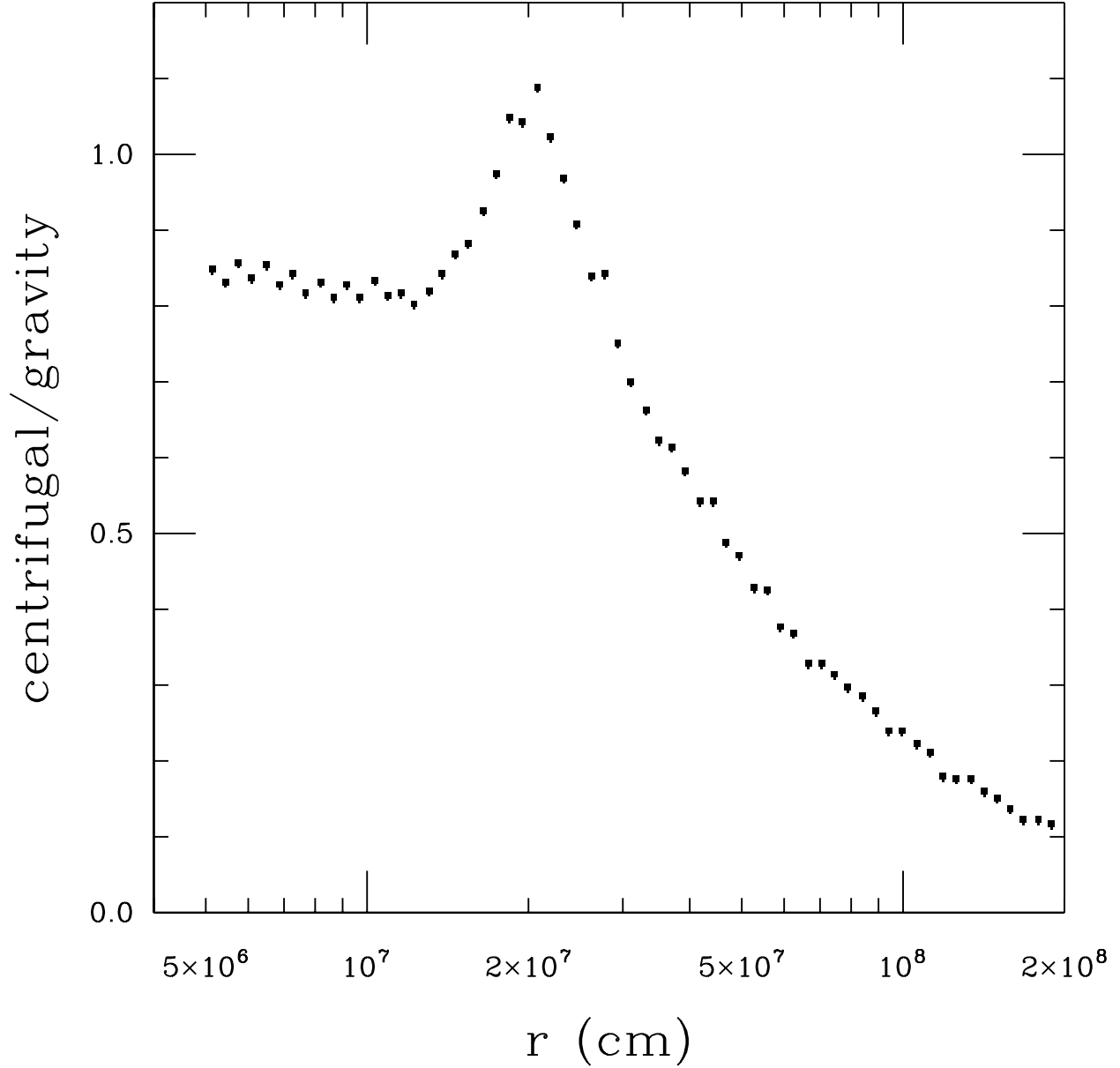,angle=90}
\caption{\label{forcebal} The ratio of centrifugal force to gravity in
Model 14A at a time 7.39 s after core collapse. An accretion shock is
located at about 200 km where a disk forms. Matter initially
overshoots the radius for centrifugal support and the ratio exceeds
unity. In the inner disk the ratio is about 0.85. The remaining 15\% is
supplied by pressure.}
\end{figure}

Fig. \ref{forcebal} shows the ratio of centrifugal force to gravity at
the same time, 7.60 s, as the other plots. At 200 km a centrifugal
barrier is encountered which is overshot by inertia. The pile up
creates the torus seen in the density plot in
Fig. \ref{popham}. Interior to 200 km, the ratio is approximately
0.85, the difference being the contribution of the radial pressure
gradient (and inertial terms).

The density structure at about the same time is given for the disk and
immediate surroundings in Fig. \ref{dens14a.small}. The maximum
density - on the grid - occurs at 200 km (larger densities exist in
the disk interior to our inner boundary) again showing the toroidal
structure of the disk and a ``pile up'' effect of the infalling
matter. Unlike the PWF calculation, our two dimensional study can
resolve vertical density structure.

At 7.60 s the density along the polar axis is already three orders of
magnitude less than in the disk.  Polar accretion occurs out to polar angles
of 30 degrees, but supersonic flow is limited to $\sim 10$ degrees. The
rotational velocity in the disk near the inner boundary is also about $c/2$.
In the polar column, temperature rises to $5 - 6 \times 10^9$ K at a density
of $5 \times 10^6$ g cm$^{-3}$. Implosive heating would lead to oxygen
burning, but probably not silicon burning. In the disk however, temperatures
are so high ($T \gtrsim 10^{10} K$) that, as previously noted, the
composition is free neutrons and protons.  In the polar region the neutrino
luminosity is $\sim 10^{22}$ erg cm$^{-3}$ s$^{-1}$; in the disk near our
inner boundary it is $\sim 10^{30}$ erg cm$^{-3}$ s$^{-1}$. Viscous
dissipation in the disk is giving a few $10^{30}$ erg cm$^{-3}$ s$^{-1}$,
several times the rate at which neutrinos can carry it away. The disk is
advection dominated.

At the same time, a plot of Mach number (Fig. \ref{mach}) shows very
supersonic accretion flow (Mach number greater than 10) along the pole
and the existence of two accretion shocks (1200 km and 200 km) in the
equator modulating the flow into the inner disk. Supersonic outflow
exists at intermediate angles due to viscous heating in the disk
($\S$4.1.5).

\begin{figure} 
\epsfig{file=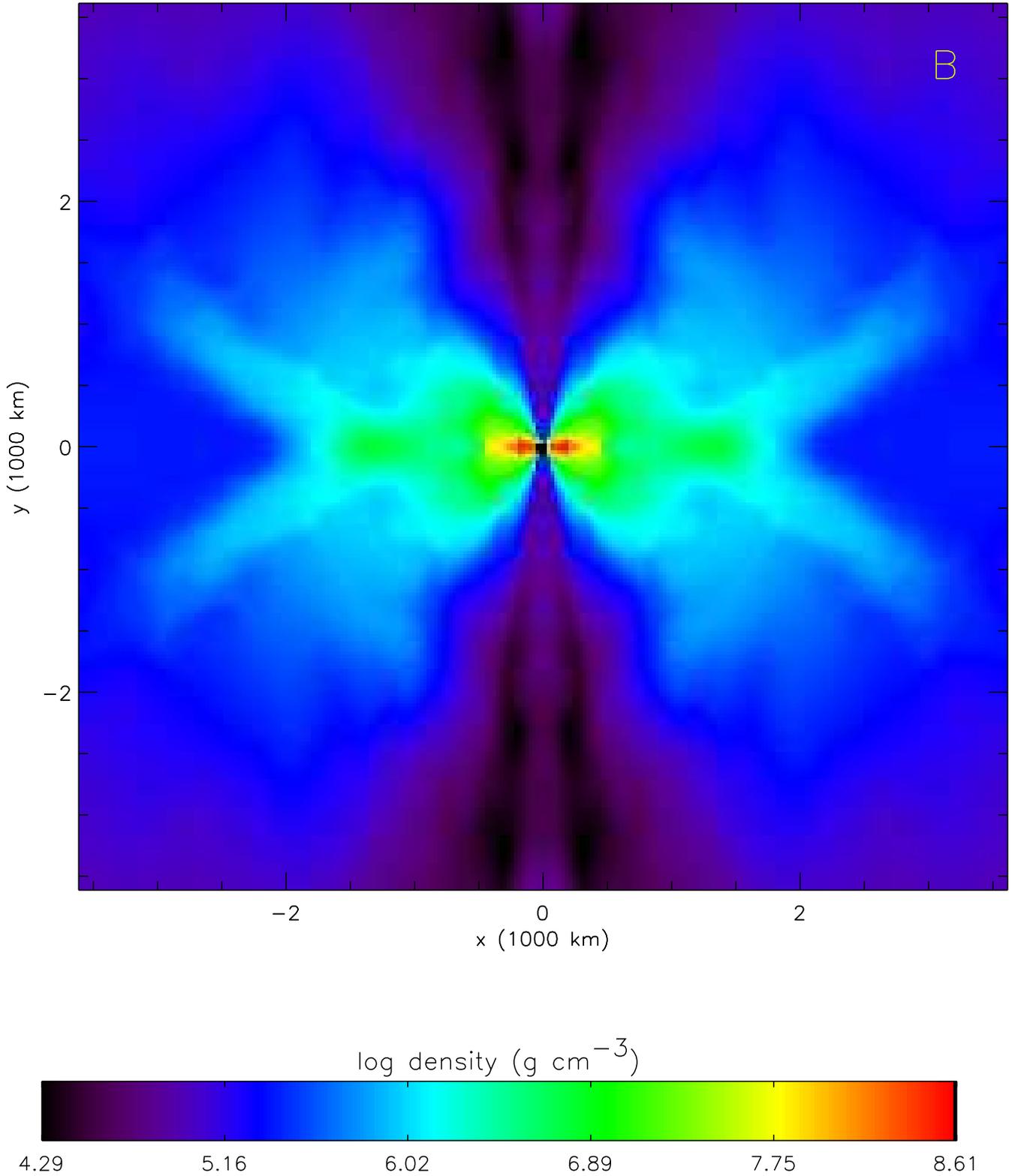}
\caption{\label{dens14a.big} A larger view of the density structure in
Model 14A at a later time (in a calculation where no jet was
initiated), 15.63 s. The inner disk with its higher density
(Fig. \ref{dens14a.small}) is unresolved here. The density along the
polar axis has declined to 10$^4$ g cm$^{-3}$, approximately the mean
density of the star.}
\end{figure}

If energy deposition from neutrino annihilation is neglected the polar
region continues to fall in and become more evaculated. Fig.
\ref{dens14a.big} shows the density structure at a late time (15.63 s) and
on a larger scale (4000 km is about 10\% of the entire star). The inner disk
is not resolved in this plot but the large density contrast between pole and
equator is still apparent and extends to large scales.

\subsubsection{The mass accretion rate}

Initially the hole accretes rapidly at all angles as the star
collapses through the spherical inner boundary.  After roughly 2 - 3 s
though, the disk has formed and it makes some sense to speak of a
disk accretion rate. Still one must continue to follow separately
the accretion that occurs along the rotational axes and that which
comes in through the disk. In practice, for Model 14A, accretion from
angles less than 45 degrees above and below the equator can be
considered ``disk-fed''.

Fig. \ref{mdot} gives that disk accretion rate for the entire duration of
Model 14A. The average rate from 5 to 15 s is about 0.07 M\sun \ s$^{-1}$,
but there is rapid time variability with episodes of accretion as low as
0.04 M\sun \ s$^{-1}$ and as high as 0.12 M\sun s$^{-1}$. Fig. \ref{mdot}
shows an expanded version of one of the enhanced accretion events.
The dots on the figure indicate a spacing of 1000 time steps. Despite the
ragged appearance of the long duration plot, the temporal structure is very
well resolved on the computer (although the accretion rate, even in the
expanded version, was only sampled every hundred time steps, i.e., about once
per millisecond).

The angular dependence of the accretion rate is shown in
Fig. \ref{mdottheta} during the same transient high value as in
Fig. \ref{mdot}. The near agreement of the rates for 45 degrees and 90
degrees (i.e., the {\sl total} accretion rate) shows that over 90\% of
the accretion is occurring thru the disk. However the disk does have
some thickness as the different value for 22.5 degrees indicates.

\subsubsection {Accretion flows and time variability}

In order to better understand the nature of the accretion
and its temporal variability, the sequence of models calculated during the
onset of the mass accretion spike near 7.60 s was singled out for careful
study.  During this spike, the mass accretion rate more then doubled in 58 ms
from 0.055 M\sun \ s$^{-1}$ at $t= 7.540$ s to 0.12 M\sun \ s$^{-1}$ at
$t=7.598$ s.  This interval was covered by over 5000 time steps in the
simulation.

\begin{figure} 
\epsfxsize=12.5 true cm
\epsfig{file=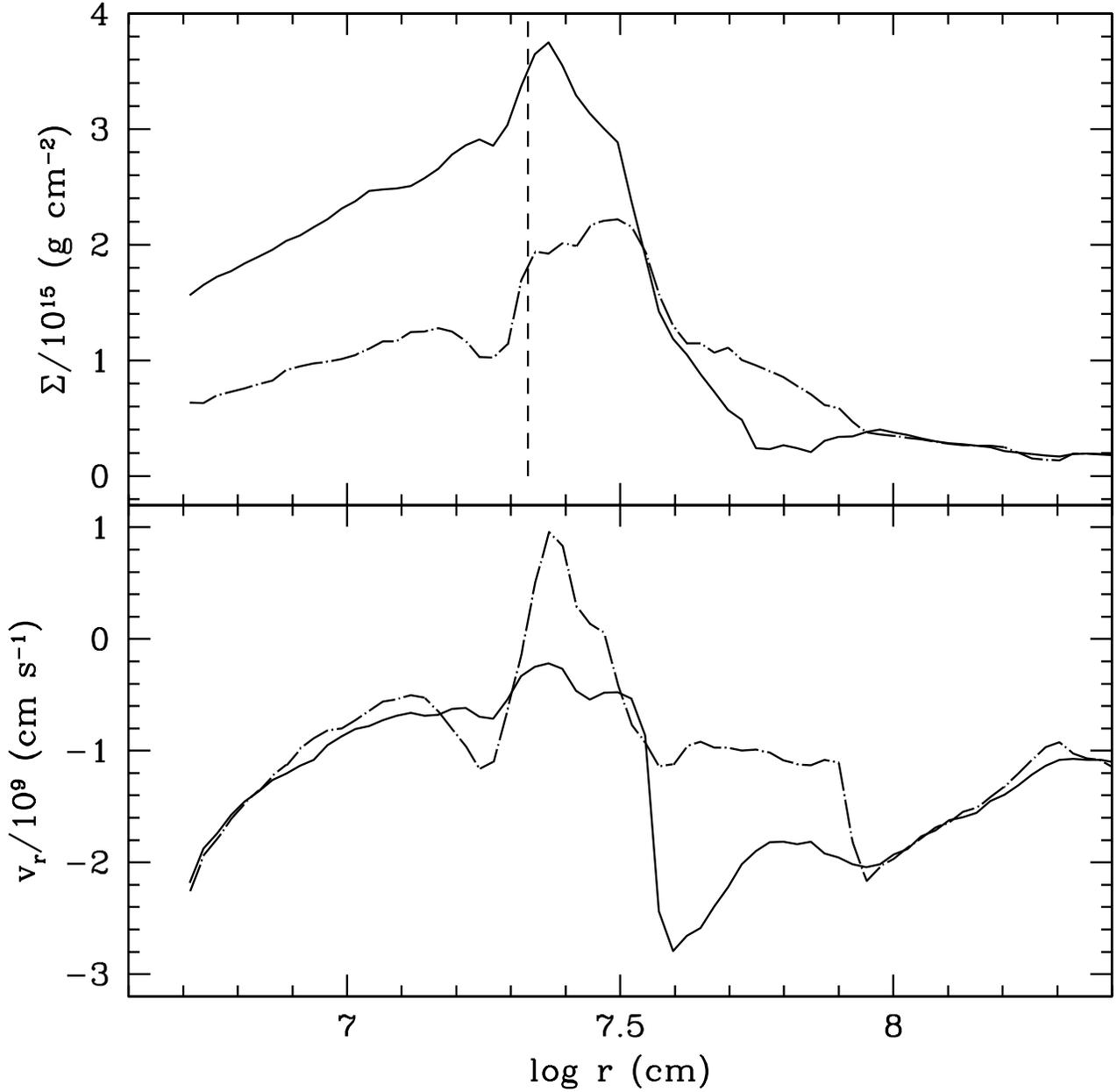,width=7 in}
\caption{\label{sigvofr} On a smaller scale the surface mass density
($\Sigma$, top panel) and radial velocity in the equatorial plane
(lower panel) as a function of radius in the equatorial plane at two
different times. The dash-dotted line is for 7.540 s, the solid line
for 7.598 s. The dashed vertical line is where gas with $j_{16}$ = 10
is supported by Keplerian rotation. The velocity in the inner disk
stays roughly constant while the density during a high accretion
episode increases due to the higher rate it is fed mass through the
shock.  Fig. \ref{massflux} shows the mass flow patern at these two
times.}
\end{figure}

\begin{figure} 
\vskip 2. true in
\epsfxsize=11.5 true cm
%\epsffile[-60 -60 410 410]{massfluxa.eps}
\epsffile[-60 -60 410 410]{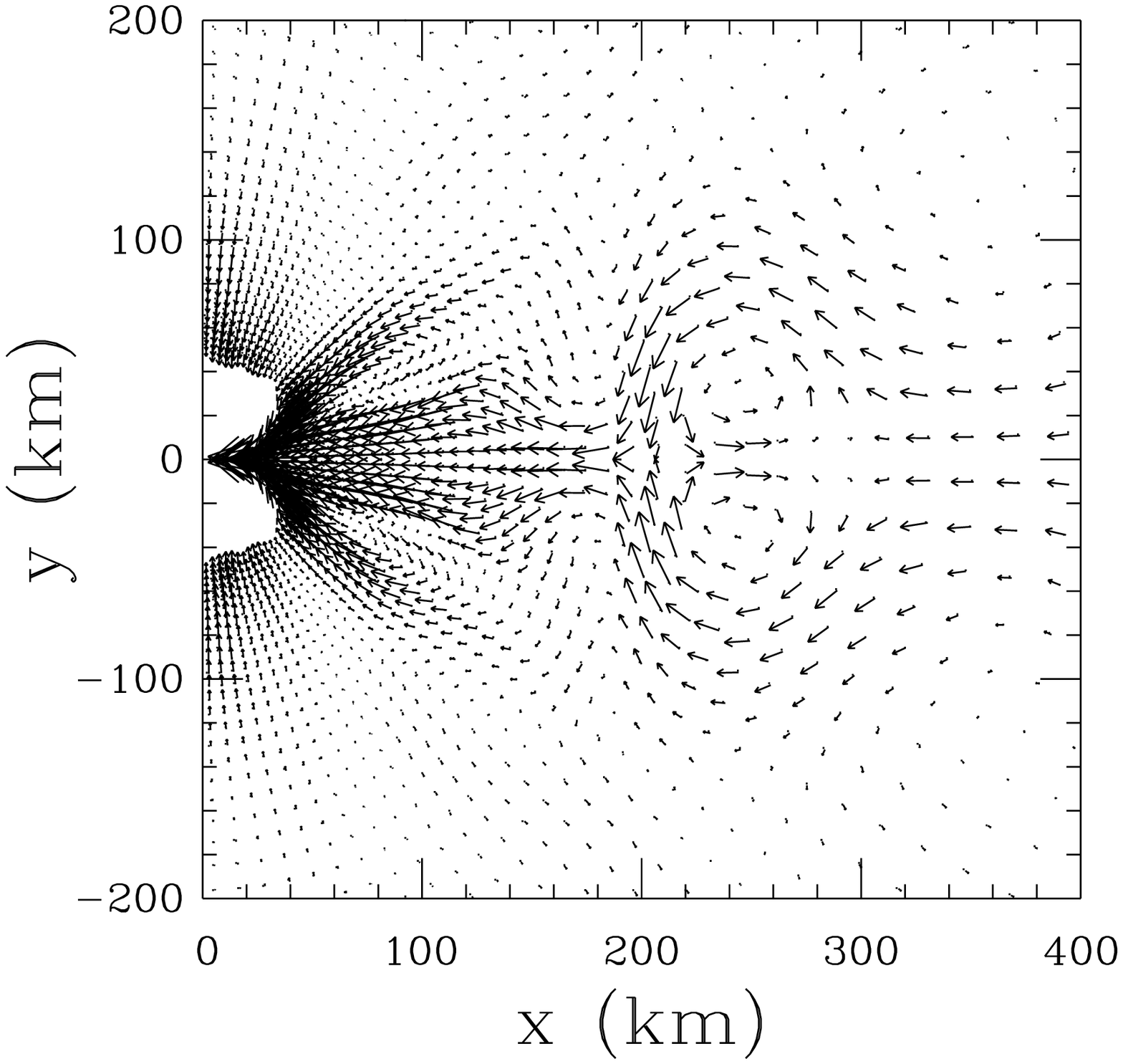}
\epsfxsize=11.5 true cm
%\epsffile[-60 -60 410 410]{massfluxb.eps}
\epsffile[-60 -60 410 410]{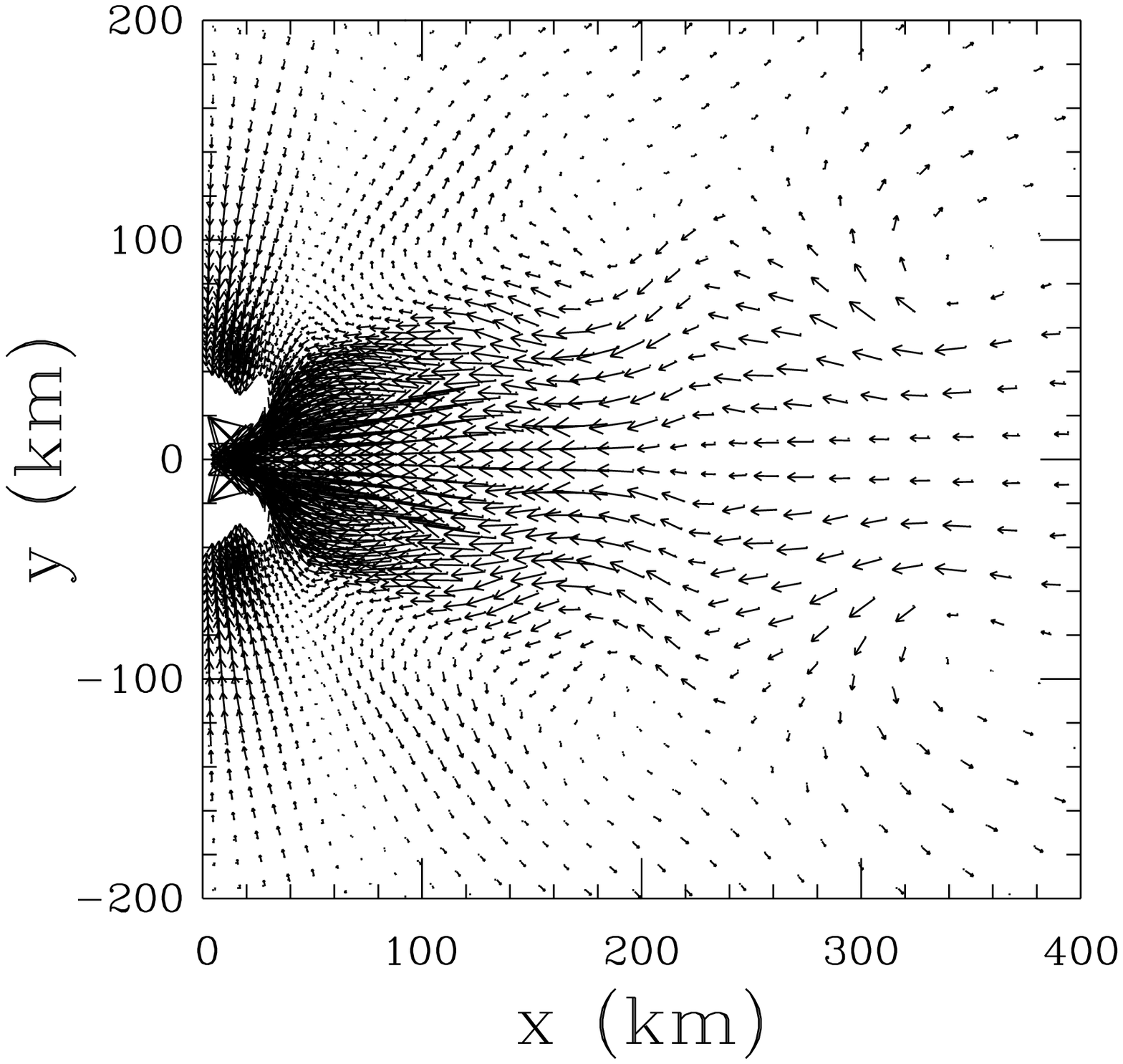}
\vskip -2.3 in
\caption{\label{massflux} The mass flux field, $\rho$ times {\bf v} at
the same times as Fig. \ref{sigvofr}. The top panel is at 7.540 s; the
lower at 7.598 s. The largest arrows are for $\rho \mathbf{v}$ $ \sim -8
\times 10^{17}$ g cm$^{-2}$ s$^{-1}$. During a high accretion state
the gate is open, the disk briefly collapses and matter flows
in. During a low state the disk grows and flow stagnates as matter
swirls at the outer boundary of the disk.} 
\end{figure}

Figs. \ref{sigvofr} and \ref{massflux} show the surface density
(density integrated along one density scale height), radial velocity, and
nucleon mass fraction in the equatorial plane, and accretion flows during the
``low'' (7.540 s) and ``high'' (7.598 s) accretion states.  Since the surface
density and velocity external to 1000 km are both constant, the accretion
rate at that radius is constant. The modulation is occurring interior to 1000
km. Fig. \ref{mdotofr}, evaluated at these same two times, shows a phase lag
in the accretion rate. When the accretion rate is
high in the inner disk, it is low outside of 400 km and vice versa.

The disk apparently has an unsteady boundary.  The location of the accretion
shock moves from 800 km (low state) to 400 km (high). As the shock moves in,
the velocity just inside the shock changes from positive to negative and the
surface density, which had been increasing, spills into the inner disk.
Thus the high accretion rate is a result of - or at least correlated
with - the collapse of the accretion shock. The time scale for this happening
is roughly the mass of the disk, $\sim$0.003 M\sun, divided by the accretion
rate, $\sim$0.1 M\sun \ s$^{-1} \approx$ 30 ms. This is also the radial
diffusion time scale for the disk and is obviously viscosity dependent.

But why should the location of the shock be unstable? We believe that
it is because of photodisintegration. In the low state, the
temperature just behind the shock is not sufficient to cause total
photodisintegration. Intact matter accumulates. But at some point
enough piles up that photodisintegration happens and, since the disk
is partly supported by pressure (but mostly centrifugal) forces, the
disk collapses into a state where photodisintegration is
essentially coincident with the shock. The disk seems to make
sporadic, irregular transitions between these two states.  However a
Fourier analysis deconvolution of the accretion rate
(Fig. \ref{fourier}) shows significant power at 50 and 25 ms. This
is approximately the viscous time scale of the disk and its first
overtone. There is also significant power at other frequencies. The
mass of the disk, the mass of the hole, the accretion rate, even the
viscosity as formulated here are all changing with time. But the fact
that significant power exists on a disk diffusion time is suggestive,
if not proof, of a real physical instability at work. Clearly this is
a subject that needs further study.

\subsubsection{Viscous induced outflows}

\begin{figure} 
\epsfxsize=12.5 true cm 
\epsfig{file=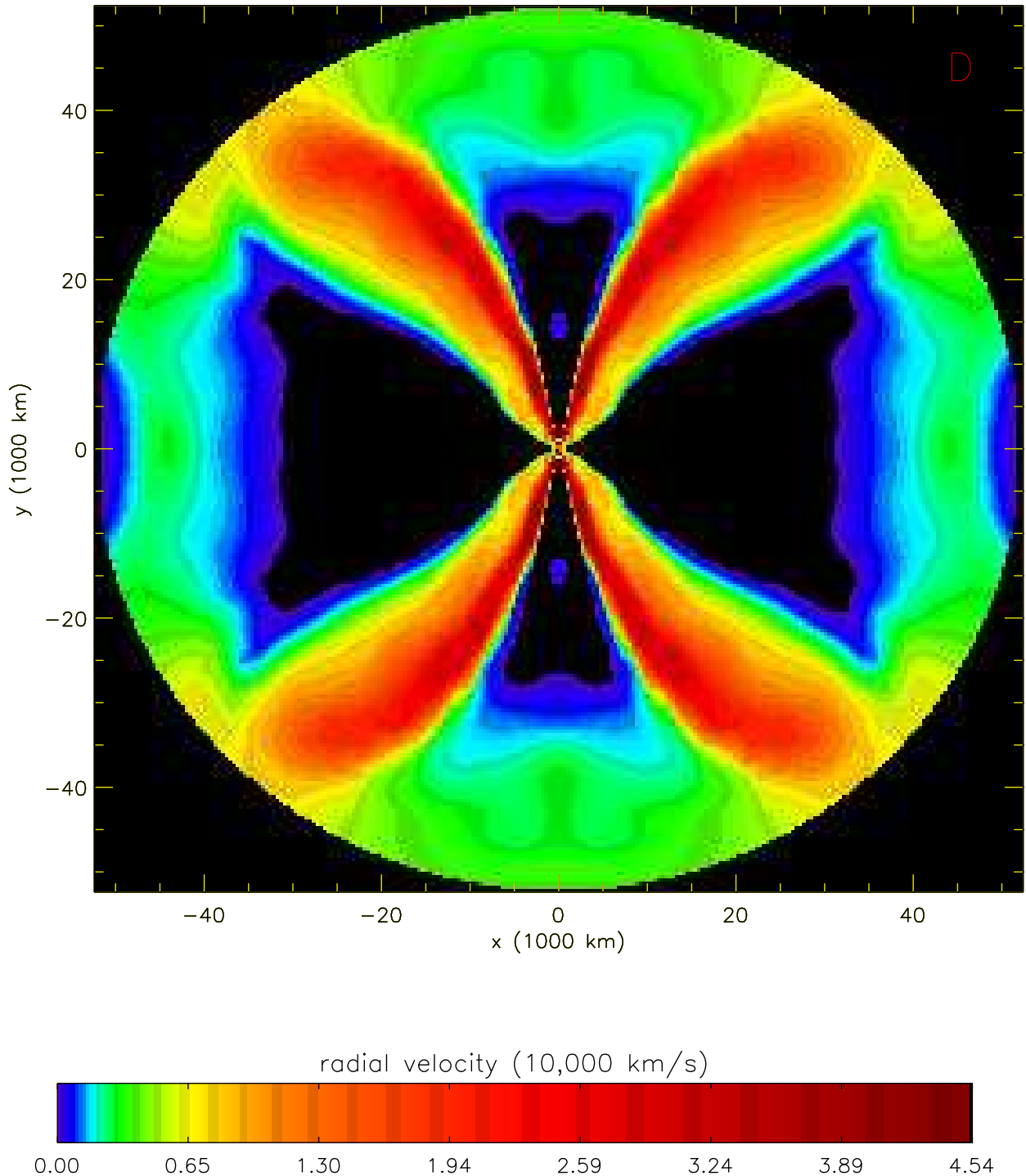,width=6.5 in}
\caption{\label{plumes} Flow patterns in Model 14A at 9.48 s on a larger
scale (50000 km). This figure shows, for a calculation in which no
neutrino energy deposition was included, powerful outflows still
developing with speeds exceeding 40,000 km s$^{-1}$ and moving outwards at
approximately 15 to 40 degrees off axis. As they approach the surface
(not shown) the velocity increases and mildly relativistic mass
ejection may occur.  These outflows are absent in Model 14B with a low
disk viscosity.  The black regions are still accreting and continue to
feed the accretion disk and black hole.}
\end{figure}

\begin{figure}
\epsfxsize=12.5 true cm
\epsfig{file=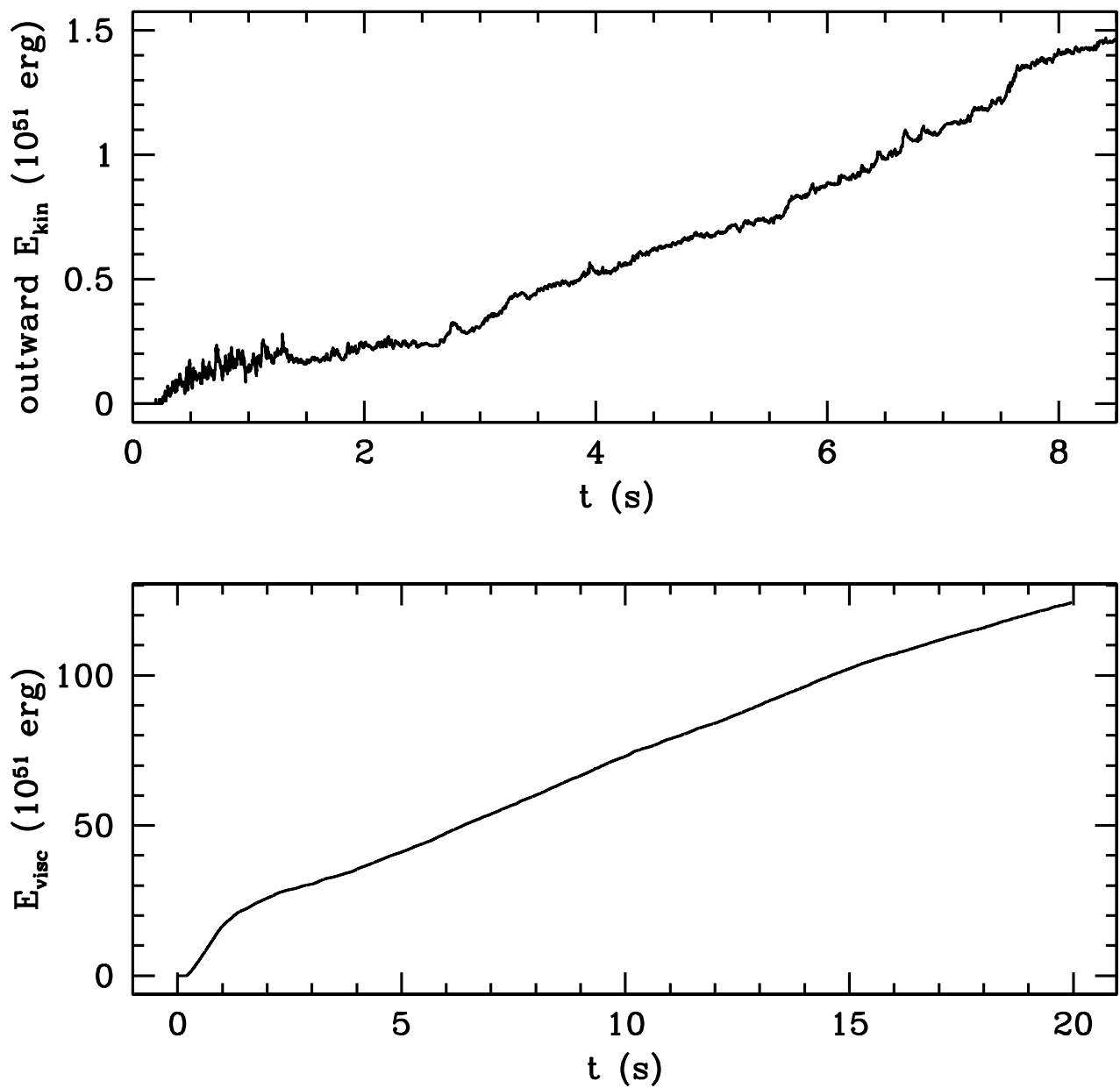,width=7 in}
\caption{\label{plumee}
Upper panel: The total outward directed kinetic energy in the
``plumes" (Fig. \ref{plumes}) grows with time. After 8 sec, material
begins to leave the computational grid, but a total energy of 2 - 3
$\times 10^{51}$ erg is estimated to have been generated in 20
sec. Lower panel: Total energy dissipated in the disk exterior to 50
km integrated over time. A much larger amount of energy is dissipated
interior to 50 km but is not included here. The energy in the plumes
is about 2\% of that dissipated in the disk on the grid carried.}
\end{figure}

While the dominant flows are polar and disk accretion, there are also
significant outflows. Viewed on a larger scale Fig. \ref{plumes} shows
plumes moving out at polar angles $\ltaprx 45$ degrees.  These flows
are only present in calculations where the disk has appreciable
viscosity (see $\S$4.2 for a low viscosity case where these outflows
are absent). They originate at $\sim$100 km as material as material
between and and two density scale heights above the disk is heated by
viscous interaction resulting from a large rotational shear.  The
entropy of this material rises (Fig.  \ref{plumeform}) and a wind is
driven off the disk. The path of the ejected mass is highly
constrained. The equator is blocked by the disk and the polar regions
by a transverse accretion shock.  The outflow follows the path of
least resistance along the outer boundary of the shock.  Over a period
of roughly 15 s, several times 10$^{51}$ erg are deposited in these
outflows (Fig. \ref{plumee}). This is roughly 2\% of the energy
dissipated on our grid. A significant portion of this energy comes
from nuclear recombination.

However, the energy in these plumes is quite sensitive to how the
(artificial) disk viscosity is treated and, in particular, to the
value of $\alpha$ adopted in the regions where the heating occurs. The
figure shown is for a calculation where the viscosity was calculated
using $\nu = \alpha \, c_{\rm s} \,r$. Another calculation, which
assumed perhaps more correctly that $\nu = \alpha \, c_{\rm s} \, H$,
with $H$, the density scale height gave about half as much energy to
the plumes. In practice this amounts to using a larger value of
$\alpha \approx 0.3$ in the latter expression. 

The plumes (or wind) are thus artificial in the sense that they are
generated by an ``artificial viscosity''. But the dissipation modeled
by $\alpha$ may have a real physical origin - magnetic energy
dissipation in and above the disk. Very roughly the MHD flux from the
disk is a small fraction, say 1 - 10\%, of the magnetic energy density
in the disk, $B^2/8\pi$, times the Alfven speed, about the speed of
light in the inner disk. The field itself might have an energy density
10\% 
of $\rho v^2$. Then for density $\sim 10^{10}$ g cm$^{-3}$, 
$v \sim 10^{10}$ cm s$^{-1}$, and a disk area of 10$^{13}$ cm, the MHD
energy input is $\sim10^{51}$ erg s$^{-1}$.

The matter that is ejected has mostly been at high temperature,$T_9
\gtaprx 10$, and is initially composed of nucleons. As these nucleons
reassemble in nuclear statistical equilibrium, and provided $Y_e$
remains near 0.5, the freeze-out composition will be dominantly
$^{56}$Ni.  The flows approach the surface with speeds in excess of
30,000 km s$^{-1}$ and may be very important in understanding the SN
1998bw phenomenon.
The accretion disk is not disrupted by these flows. Accretion
continues even as the star blows up at angles above $\sim$45
degrees. Several solar masses remain at this point 
outside the disk in the equatorial plane.

\subsubsection {The evolution of the Hole Mass and Kerr parameter}

As the black hole accretes, both its mass and angular momentum
grow. The hole might be born without rotation, but more realistically,
had some initial angular momentum, that is a normalized Kerr
parameter, $a \equiv Jc/GM^2$, that was significantly greater than
zero.  The angular momentum in the iron core of the presupernova star
(Fig. \ref{heger}) corresponds to $a_{\rm init} \approx 0.5$. This is
also a reasonable value if the black hole forms from a contracting
proto-neutron star born at near break up. In what follows, we will
consider both $a_{\rm init} = 0$ and $a_{\rm init}$ = 0.5 as
interesting cases, though a$_{\rm init}$ = 0.5 is a choice more
consistent with the angular momentum distribution assumed for the
mantle.

The initial gravitational mass of the black hole is also relevant; a
lighter hole can be spun up more easily. The contracting protoneutron
star that made the black hole radiated some portion, as much as
$\sim$30\%, 
of its rest mass as neutrinos before the collapse became
dynamic. This fraction is uncertain and depends on the entropy of the
neutron star as well as the EOS. Here we make the conservative (for
making a GRB) assumption that the black hole mass is initially the
baryon mass  of the iron core that is removed - 2.03 M\sun. 

The material that falls into the hole mostly accretes in the
equatorial plane from a nearly Keplerian orbit. The contribution of
pressure in the inner disk is about 15\% (Fig. \ref{forcebal}) and
even material coming in from high latitude has so much angular
momentum that it cannot go directly into the hole. The inner disk is
also slim (Fig. \ref{popham}).

Since we have not carried the inner disk on our numerical grid,
analytic models have to be used to specify the rate at which
gravitational mass and angular momentum are added for a given baryonic
accretion rate. Because the results are sensitive to them, the
evolution of the Kerr parameter and gravitational mass were computed
using three different models to extend our inner boundary at 50 km
($\approx 10 r_g$, where $r_g \equiv GM/c^2 = 4.5 (M/3M_{\odot})$ km)
to the event horizon at 2$r_g$ for a Schwarzschild hole ($a=0$) and at
$r_g$ for an extreme Kerr hole ($a=1$).  Fig. \ref{mbha} shows the
evolution of $a$ and $M$ for the limiting assumptions of 1) a ``thin''
disk in which all heat generated by viscosity and compression is
assumed to be radiated away (Bardeen, 1970); 2) an advection dominated
accretion disk (ADAF) in which no heat escapes,all is advected into
the hole with the accreting gas (Popham \& Gammie, 1998); and 3) the
intermediate case of a neutrino dominated disk which radiates part of
its heat and mass as determined by the appropriate neutrino emission
processes (PWF). Presumably this last case is the most realistic.

For the ADAF and neutrino dominated disks, the gravitational mass and
angular momentum of the black hole were calculated by multiplying the
mass accreted through our inner boundary each time step by the energy
per unit rest mass ($e$) and angular momentum per unit rest mass ($j$)
(both functions of the Kerr parameter a) at the event horizon.  For
the ``thin'' disk, the values for $e$ and $j$ at the last stable
circular orbit were used. This makes the assumption that the disk
interior to our inner boundary is capable of transporting the mass
delivered to it to the event horizon.  The fact that the disk interior
to about 200 km does smoothly transport the mass it receives from the
star gives us confidence that this is a good assumption.  For the hole
mass, the accretion through all polar angles was used while for the
angular momentum only mass accreted within $45^{\circ}$ of the equator
was used.  In practice, this choice has very little effect on the
calculated quantities since most mass is accreted within $45^{\circ}$
of the equator.

For the thin disk, the energy and angular momentum at the last stable
circular orbit ($r_{lso}$) are given by Bardeen (1970) and Bardeen,
Press, \& Teukolsky (1972) $$j_{\rm lso} \ = \ {{2 G M_{\rm bh}} \over
{3^{3/2}
\,c}} \, \left(1
\, + \, 2(3 \,z \, - \, 2)^{1/2} \right)$$
where $z \equiv r_{\rm lso}/ r_{\rm g}$, $M_{\rm bh}$ is the
gravitational mass of the black hole and $r_{\rm g}$ is the
gravitational radius, $G M_{\rm bh}/c^2$, and $$r_{lso}=M
(3+Z_2-((3-Z_1)(3+Z_1+2Z_2))^{1/2})$$ with $$Z_1 = 1 + (1 -
a^2)^{1/3}((1 + a)^{1/3} + (1-a)^{1/3})$$ and $$Z_2 = (3a^2 +
Z_{1}^{2})^{1/2}. $$ The corresponding energy per unit mass at the
last stable orbit is 
$$e_{\rm lso} \ = \ \left[ 1 \, - \, {{2} \over
{3 z}} \right]^{1/2}
\, c^2. $$
Because the ``thin'' disk rotates at the maximum (Keplerian)
rate and radiates away the entire binding energy of the disk gas it
results in the highest $a$ and the lowest $M_{\rm bh}$.

For the ADAF, we used the specific angular momentum at the event
horizon of Popham \& Gammie (1998) ($\alpha = 0.1$) while the energy
per unit mass is simply $c^2$ since the entire mass-energy of the gas
is assumed to accrete into the hole.  The ADAF solution produces the
largest hole mass and lowest $a$ because the full rest mass is
accreted (no energy escapes) and because non-negligible radial
pressure gradients in the disk result in significantly sub-Keplerian
(by as much as 27\%) angular momentum in the accreting material.

The values of $j$ and $e$ at the event horizon for the neutrino dominated
disk were interpolated in $a$ from the $M_{bh} = 3 \ M_{\odot}$ ,
$\dot{M}=0.1 M_{\odot}$ s$^{-1}$, $\alpha = 0.1$ results of PWF and were
provided by Popham (private communication).  Since these disks only radiate a
fraction of their binding energy, they are intermediate between the ADAF and
the ``thin disk''.  The ``standard'' disk of PWF radiates less than half its
binding energy at $a=0$, but the radiative efficiency increases with $a$ and
the inner disk becomes thinner.  For $a > 0.9$ the disk radiates its binding
energy efficiently and therefore produces $a$ and $M$ similar to the
``thin'' disk.

As $a$ approaches unity the rotational energy in
the hole also becomes enormous $E_{\rm rot} \approx 10^{54}$
erg. Extraction of even a small fraction of this energy by MHD
processes (e.g., Blandford \& Znajek 1977; Katz 1994, 1997;
M\'esz\'aros \& Rees 1997) will dominate even over the large energies
we now compute for neutrino mediated energy transport.

\subsubsection {Estimated neutrino luminosity and energy deposition}

\begin{figure} 
\epsfxsize=12.5 true cm
\epsfig{file=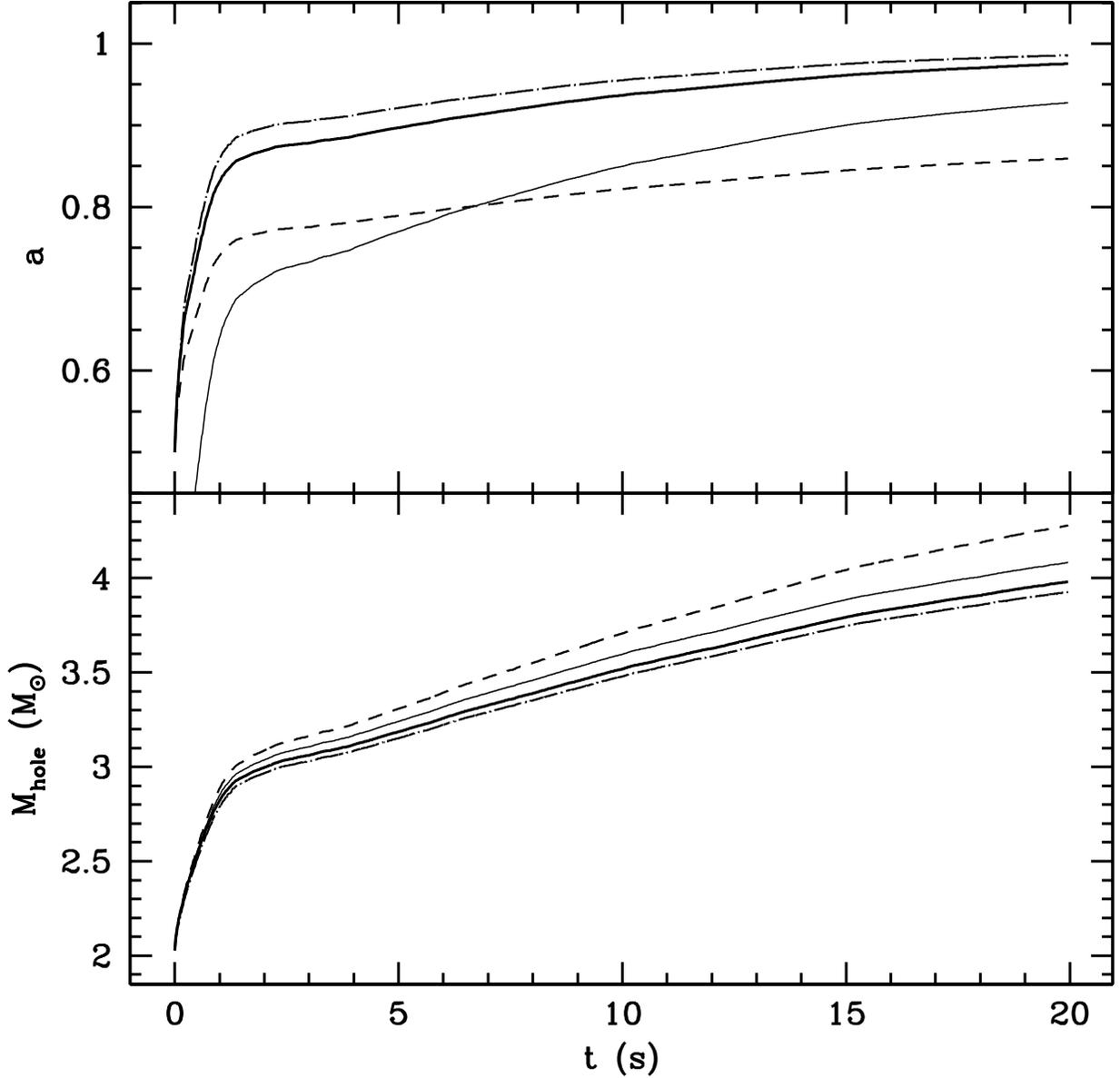,width=6.75 in}
\caption{\label{mbha} The hole mass and normalized Kerr parameter grow
rapidly in the first second as the dense stellar core collapses
through the inner boundary at all polar angles. After the first solar
mass is accreted (in approximately one second) centrifugal forces begin
to halt the collapse along the equator and an accretion disk
forms. The upper panel shows the increase in the
Kerr parameter for various models for the disk interior to the inner
boundary at 50 km. ``Thin'' (dash-dot), neutrino-dominated (thick solid)
and advection dominated (short dash) models are shown for initial Kerr
parameter a$_{\rm init}= .5$.  The thin solid line shows the neutrino
dominated case for a$_{\rm init} = 0$.  The value of $a$ at 20 s for the four
lines given is .9849, .9752, .8591, and .9274.  The lower panel shows
the growth of the gravitational mass of the black hole.  The
short-dashed line shows the growth in baryonic mass of the black hole
since for a pure advective model no energy escapes the inner disc.}
\end{figure}

One of the useful implications of the good agreement of our model with
that of PWF, in the inner regions where the disk is steady state
(Fig. \ref{popham}), is that their solution can be used to extrapolate
our own to the event horizon. In particular, we can use their
estimates of {\sl total} neutrino luminosity (as opposed to the small
fraction we calculate on our grid, e.g., Fig. \ref{plumee}), and their
neutrino annihilation efficiency as a function of accretion rate, Kerr
parameter, and black hole mass. In what follows we assume a disk
viscosity parameter $\alpha$ = 0.1. Extrapolation to other values
follows using PWF. Because the PWF tables are sparse for black hole
masses other than 3 M\sun, because the neutrino luminosities and
efficiencies are not rapidly varying with hole mass, and because our
black hole mass stays at all times within 50\% of 3 M\sun, we also
assume a constant value of black hole mass equal to 3 M\sun \
(Fig. \ref{mbha}).

PWF showed that the efficiency for neutrino emission and energy
deposition along the rotational axes is very sensitive to both the
accretion rate and the Kerr parameter. Table 1, extracted from their
work and amended by additional calculations performed by Popham \&
Fryer specifically for the collapsar model, gives some key quantities
for our range of accretion rates and Kerr parameters.
 
One sees, as also noted by PWF, a transition in disk
behavior for $\dot M \ltaprx 0.05$ M\sun \ s$^{-1}$. For lower
accretion rates, the disk is increasingly advective. Energy dissipated
in the disk is carried into the hole and not effectively radiated in
neutrinos. At higher accretion rates, both the neutrino luminosity and
the efficiency for neutrino annihilation increase sharply. The
efficiency depends quadratically on the luminosity and also on the
neutrino temperature, both of which are higher in the high $\dot M$
case. These quantities also increase very sharply with Kerr parameter,
$a$. As $a$ becomes larger (Fig. \ref{mbha}), the last stable orbit
moves in. Emission from the higher gravitational potential increases
both the luminosity and the temperature and also makes the density of
neutrinos higher because of the more compact geometry.

A full calculation of the neutrino transport in the situation
considered here is a formidable problem, rivaling, perhaps
exceeding that of a neutrino powered supernova of the ordinary
variety. For the most part, the disk is optically thin, but especially
in the high Kerr parameter cases of greatest interest, it is becoming
grey. This violates one of the key assumptions of PWF. Also the
trajectories of the neutrinos are not straight lines (see Ruffert \&
Janka 1998), but follow geodesics. We have had to make a number of
approximations to translate our mass accretion histories into energy
deposition efficiencies.

First, red shifts are included in the PWF calculations by keeping
track of the gravitational potentials where neutrinos are emitted and
absorbed. But the neutrinos are assumed to go in straight lines, not
follow geodesics. This may not be too bad an assumption (Ruffert \&
Janka 1998) for $a = 0$, but becomes increasingly suspect at small
radii as $a$ approaches 1. For the particular study here, Popham and
Fryer calculated two cases for each value of accretion rate and Kerr
parameter. We shall refer to these as the ``conservative'' and
``optimistic'' cases - though various people may have different views
regarding these terms. For the conservative case, all neutrino
emission and all neutrino annihilation are neglected if the emission
or absorption occurs within two event horizon radii of the
origin. This makes little difference for the $a = 0$ case, but for $a
= 0.95$, the reduction in energy deposition can be appreciable (Table 1).
Neutrino annihilation within one scale height of the disk is also neglected.
Further, in any region where the neutrinos might be considered ``trapped'',
the flux out of the disk is set to zero. This trapping decision is made based
upon the neutrino diffusion time, as determined by the disk thickness,
density, and temperature, compared to the local accretion time scale,
$r/v_r$.   The optimistic case makes similar assumptions about the
annihilation region - a disk scale height is excluded, a region of two event
horizons is excluded - but luminosity from neutrinos all the way down to the
last stable orbit are included. This is particularly important when the Kerr
parameter is large.

\begin{figure} 
\epsfxsize=12.5 true cm
\epsfig{file=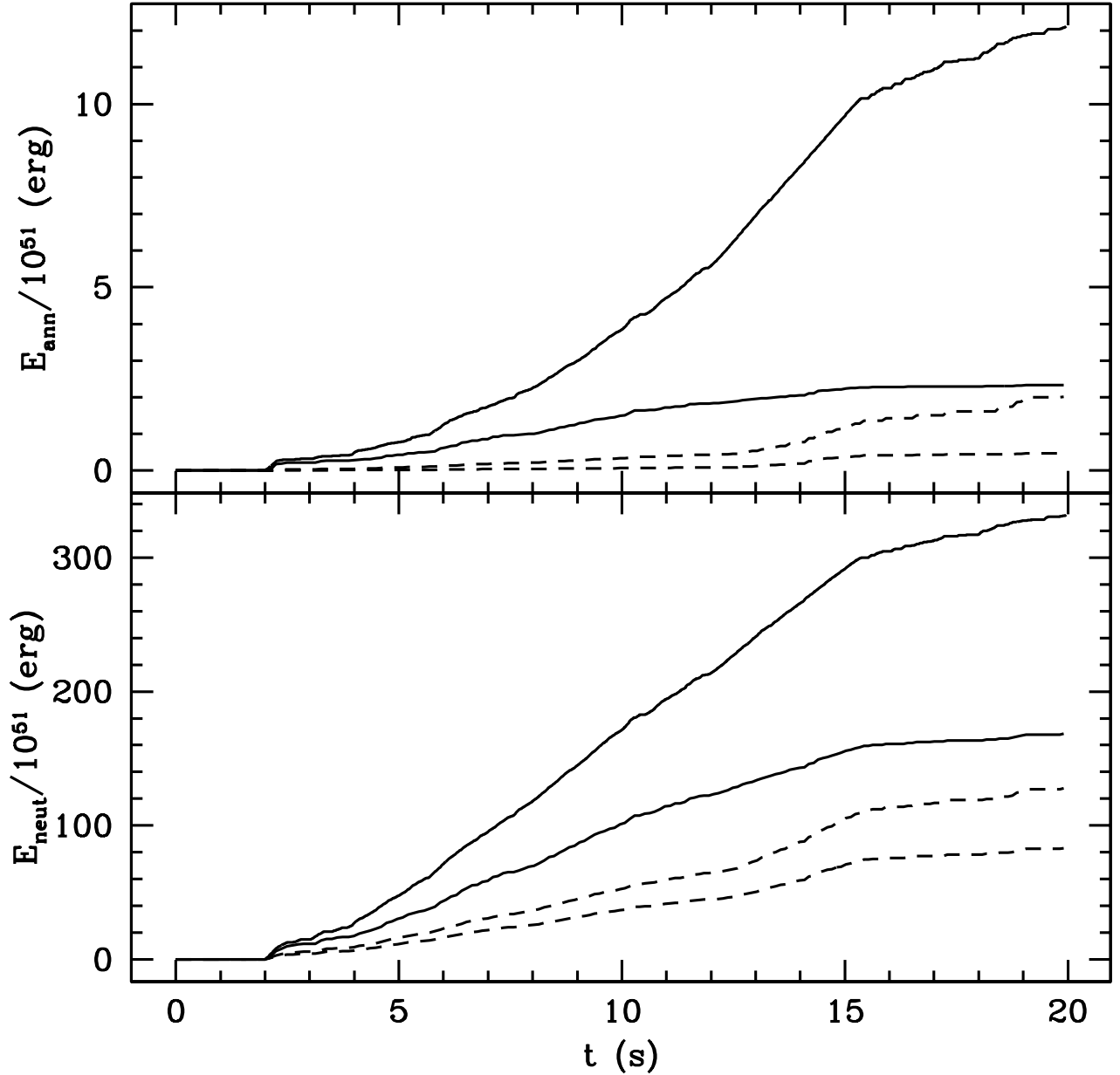,width=7 in}
\caption{\label{etot} Time integrated neutrino annihilation
energy. The top panel shows the running integral of the energy
deposited for two choices of initial Kerr parameter($a_{\rm init}$ =
0: dashed lines; $a_{\rm init} = 0.5$: solid lines) and for two
assumptions regarding the efficiency of neutrino annihilation (see
text). The higher lines for each case use the ``optimistic'' neutrino
rates.  The bottom panel gives the {\sl total} neutrino energy radiated from
the disk for the same assumptions.}
\end{figure}

We then calculated the neutrino luminosity and energy deposition by
neutrino annihilation for Model 14A by (logarithmic) interpolation in
Table 1. Consistent with our desire to set reasonable upper and lower
limits, we considered both the ``conservative'' and ``optimistic''
neutrino transport approximations and also black holes born stationary
and with $a_{\rm init}$ = 0.5. Fig. \ref{edep} shows both the neutrino
luminosity (chiefly neutrinos from pair capture on nucleons) and the
energy deposited for a typical case in which the conservative neutrino
transport scheme was employed and the initial rotation of the hole was
$a_{\rm init}$ = 0.5. Fig. \ref{etot} shows the integrated luminosity
and energy deposition for four cases. For the optimistic case, the
total energy was $12 \times 10^{51}$ erg and would clearly have been
greater had we followed the calculation further. For the most
conservative case of an initially non-rotating  black hole and
restricted neutrino transport, the energy fell to $0.5 \times 10^{51}$ erg.

We feel that the ``optimistic'' curves are realistic, but need a more
careful treatment of the neutrino physics before fully trusting
them. It should be noted however, that even the optimistic curves may
underestimate the total energy available from the collapsar model. A
sustained accretion rate above 0.1 M\sun \ s$^{-1}$ (instead of the
average here - 0.07 M\sun \ s$^{-1}$) gives a much higher
energy. Multiplying the accretion rate by 1.5 in Model 14A for example
and using $a_{\rm init}$ = 0.5 gives total energies after 20 seconds
of $4 \times 10^{51}$ erg (conservative) and $30 \times 10^{51}$ erg
(optimistic) respectively (however, the black hole mass was not
increased in a manner consistent with the higher accretion rate in
this example. It is thus a slight overestimate of the increased
efficiency). Such an increase in accretion rate might be easily
achieved in a star of higher mass, lower angular momentum, or
different disk viscosity. Decreasing the initial mass of the black
hole because of neutrino mass loss in the protoneutron star can also
raise the numbers. More massive stars might also deliver energy
$\sim$10$^{51}$ erg s$^{-1}$ for considerably longer than 20 s.

One also sees very significant variability in the neutrino energy
deposition (Fig. \ref{edep}). This is because of the time varying
accretion rate (Fig. \ref{mdot}) and the sensitivity of the neutrino
efficiency to accretion rate (Table 1).  Varying the accretion rate
from 0.05 to 0.1 M\sun \ s$^{-1}$ will change the energy deposition by
neutrino annihilation by a large factor (20 for $a$ = 0.95). Because
the light (or neutrino) crossing time in the vicinity of the hole is
about a millisecond, and because the matter in which neutrinos are
deposited moves at about c, the jet produced by neutrino energy
deposition ($\S$5) can change its energy in almost instantaneous
response to the accretion rate. One thus expects a highly variable
energy for the jet. The implications of this are explored in $\S$ 
6.1.1.

\subsection {A low viscosity model}

Model 14B was calculated in an identical fashion to Model 14A, but
employed a much smaller disk viscosity. Indeed it was initially our
intention to calculate a ``zero'' viscosity model for comparison to
$\alpha = 0.1$ in Model 14A, but we found that all hydrodynamic codes,
even PPM, have some numerical viscosity. By setting our external
viscosity parameter $\alpha$ to zero, we were able to determine 
the effective internal $\alpha$ of the code itself. By comparison to density
distributions in PWF, that value of $\alpha$ is about 0.001.

\begin{figure} 
\epsfxsize=12.5 true cm
\epsfig{file=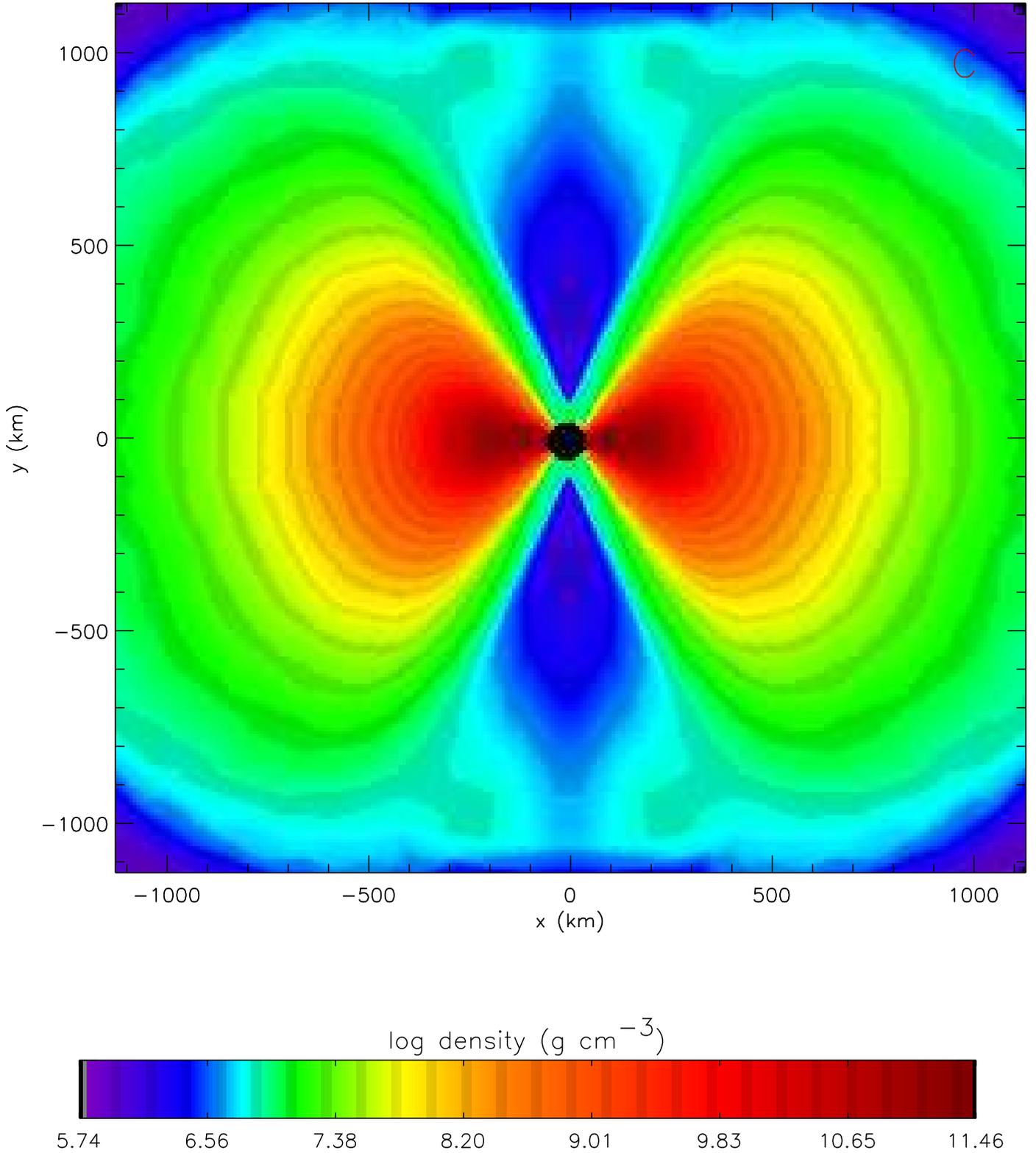}
\caption{\label{lowalpharho}
Density structure at 7.52 s in the low viscosity run, Model 14B.  Note
the higher density contrast and more massive disk than in the
calculation with high viscosity (Fig. \ref{dens14a.small}).  The
opening angle of the accretion column is also larger and viscosity
induced outflows are absent.}
\end{figure}

Figs. \ref{lowalpharho} and \ref{mdotloalp} show the density structure
and the accretion rate for this model. Because of the lower disk
viscosity and the almost identical mean accretion rate (the low
$\alpha$ run actually has a little higher $\dot M$ due to the absence
of viscosity driven outflows, $\S$4.1.5) both the density in and mass of
the accretion disk are much higher. The contrast between polar density
and equatorial density is correspondingly greater.  At 9 s, interior
to 200 km, the disk mass is 1.25 M\sun; interior to 300 km it is 2.21
M\sun. For Model 14A with high disk viscosity the corresponding
numbers were (at 20 s) 0.0016 M\sun \ and 0.0033 M\sun. For disk
masses as high as $\sim$1 M\sun \ self-gravity will become important
and gravitational instabilities (e.g. spiral arm formation) can start
to transport angular momentum in the disk.  Clumpiness due to
self-gravitating blobs in the disk may also lead to time structure in
the accretion rate.

The time history of the accretion rate is similar to that for Model
14A (Fig. \ref{mdotloalp} $vs.$ Fig. \ref{mdot}), though noisier. It
shows some time structure, but, is generally less time variable than 14A. The
disk instabilities discussed in $\S$4.1.4 are weak or absent. A
Fourier analysis of the accretion rate (not shown) yields no
characteristic frequencies.  We also see that the outflows produced in
the higher $\alpha$ model are absent in Model 14B.

The lower temperature in the disk reduces the neutrino emission and
makes it more ``advective'', much less likely to power a GRB by neutrino
energy deposition. On the other hand the disk goes through many
more revolutions before accreting and also has a much higher energy density,
$\rho v^2$. A (larger) equipartition field has more time to develop and MHD
energy extraction may be more efficient ($\S$4.1.5).

\subsection {A model with high angular momentum and nuclear burning}

Bodenheimer \& Woosley (1983) also explored a model similar to those
studied here based upon the failed explosion of a $M_{\alpha} \approx
9$ M\sun core. However they used a comparatively large value of
specific angular momentum, namely whatever value was necessary to give
centrifugal force divided by gravity = 4\% {\sl at all cylindrical
radii}. In practice this corresponded to $j_{16} \approx 5 -100$
(compare to our standard value here, $j_{16} \approx 10$). They
also used a much larger inner boundary radius (1500 km) and
experimented with a finite pressure gradient at that boundary. For
such large $j$, a centrifugal bounce and explosive oxygen burning
happen at approximately the same radius where a disk might form, a few
thousand km. A combination of nuclear burning and rotation thus gave a
weak equatorial explosion accompanied by the synthesis of some
intermediate mass elements and a little $^{56}$Ni.

\begin{figure} 
\epsfxsize=12.5 true cm
\epsfig{file=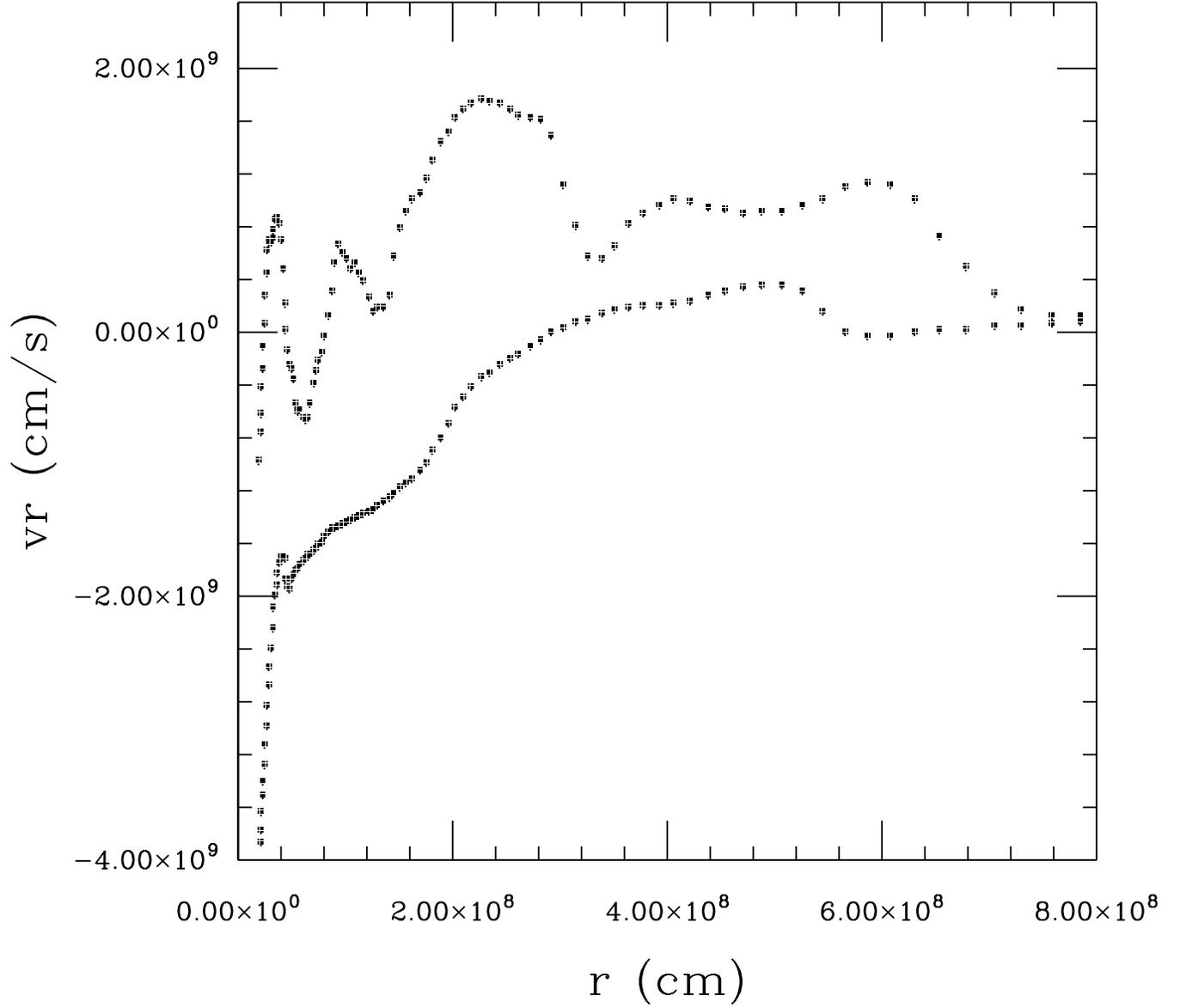,angle=90,width=7 in}
\caption{\label{vox}
Velocities in the equatorial plane for two runs that did (upper line) and
did not (lower line) include energy generation from oxygen burning.}
\end{figure}

We carried out a similar calculation here for $M_{\alpha}$ = 9 M\sun \ (i.e.,
the core of the 25 M\sun \ presupernova) - but with the inner boundary moved
in to 200 km. For this smaller radius centrifugal support completely
dominated the force balance and a zero pressure gradient boundary condition
could be used. Explosive oxygen burning and equatorial outflow were observed
in the first few seconds (Figs. \ref{oxflow} and \ref{vox}). However, since
we are mostly interested here in GRBs not weak supernovae, that calculation,
which posed some numerical difficulty, was halted at 6 s. The implication is
though that too much angular momentum, as well as too little, can keep an
optimal accretion disk from forming and inhibit the GRB phenomenon.

\section{EXPLOSIONS AND JETS}

Given the large energy deposition calculated for the standard model
($\S$4.1.5, $\S$4.1.7, and Fig. \ref{edep}) one expects an explosion of some
sort. But will the outflow be relativistic at any angle, especially
near the rotational axis, and, if so, how well collimated will it be?
Can the major reservation often voiced regarding the collapsar
model - the ``baryon loading'' - be satisfied? And will the rest of
the star explode or accrete? 

We deposited thermal energy along the rotational axes at a rate
comparable to that calculated in $\S$4.1.7, namely $5 \times 10^{50}$
erg s$^{-1}$ at each pole for a total of 10$^{51}$ erg s$^{-1}$.  This
energy was deposited by adding thermal energy uniformly in a region
bounded by 50 km to 150 km above and below the black hole for a range
of polar angles 0 to 30 degrees.

However, we did not begin this energy deposition immediately. It is
not possible to produce a strong outflow very early when the momentum
of the infalling material along the axis is too high. At $\sim 1$ s
for example, the density of the infalling material is $\gtaprx$10$^7$
g cm$^{-3}$ and its velocity, $\sim 10^{10}$ cm s$^{-1}$ corresponding
to a kinetic energy influx of $3 \times 10^{51} r_7^2 v_{10}^3$ erg
s$^{-1}$ where $r_7$ is the cross sectional radius of the accreting
region in units of 100's of km and $v_{10}$, the accretion velocity in
units of 10$^{10}$ cm s$^{-1}$.  Here $r_7$ was approximately 0.5 (the
area of our inner boundary). Unless the deposition rate is comparable
to this, any energy deposited by neutrinos or MHD processes will be
advected into the hole. As time passes however, the velocity remains
approximately constant, but the density declines. By $\sim$7 s after
black hole formation, the density has declined to $\ltaprx 10^7$ g
cm$^{-3}$ and the accretion energy to $\sim$ few $\times 10^{50}$ erg
s$^{-1}$. At this point energy deposition $5 \times 10^{50}$ erg
s$^{-1}$ becomes capable of reversing the inflow.

Starting at 7.60 s in Model 14A energy was deposited as
described above. The density in the accretion column had fallen to $1
\times 10^7$ g cm$^{-3}$, but there was a sharp decline to 10$^6$ g
cm$^{-3}$ at 200 km. It may be that explosion could have been induced
at a somewhat earlier time (but no earlier than 2 - 3 s when the disk
formed).  The choice of 7.60 s is somewhat arbitrary. Because of the
computational expense imposed by the Courant condition, we waited for a
situation where the jet could clearly begin at least an initial propagation
on our grid.  During the next 0.45 s, no outward motion developed, but the
density declined between 50 and 200 km by a factor of about 30. This led to a
decrease in the ram pressure and set the stage for a velocity reversal. Only
0.15 s later, 8.20 s, the density above the hole at all distances was less
than 10$^5$ g cm$^{-3}$ and outward velocities had developed. A shock bounded
rising material as it encountered continuing accretion. The shock at his
time was located at 2500 km and highly collimated with an opening angle of
$\sim$10 degrees (here and elsewhere the ``opening angle'' is one-half of the
total angle). This implies a shock speed during this interval of at least
20,000 km s$^{-1}$. The temperature at the base of this bubble was $8 \times
10^9$ K and the density, about 10$^4$ g cm$^{-3}$ corresponding to an energy
density of 10$^{22}$ erg g$^{-1}$. Were this material to expand freely, it
would already become relativistic with $\Gamma \sim 10$.

But the bubble cannot, at this stage, expand freely because the star is in
the way. A channel for the radiation and pairs must be cleared to the
stellar surface and this takes time. Such a clearing is possible though
because energy deposition at the base continues even as the density declines.
Neutrino annihilation depends only on geometrical factors, neutrino energy
and neutrino luminosit, not on the local density. We continued to deposit
energy at the same rate per unit volume.

It is worth noting here two shortcomings of our calculation, both of
which cause us to underestimate the momentum of the jet. First,
neutrino annihilation does not just deposit energy, it also deposits
momentum. When a neutrino from one side of the disk meets its
counterpart from the other, symmetry requires a net momentum along the
rotational axis in the outgoing electron-positron pair. Since the
collision angle is not large and all the particles involved are
relativistic, the momentum deposited is approximately the energy
deposition divided by $c$. The amount can be considerable. An
energy deposition of $5 \times 10^{51}$ erg would provide enough
momentum to move 0.1 M\sun \ at almost 10,000 km s$^{-1}$. Second,
most of the mass-energy in the jet is in the form of radiation and
pairs, yet, in our non-relativistic hydro-code, only baryons carry
momentum. So the actual directed momentum of the jet is far greater
than we calculate and it would penetrate the overlying star quicker
and easier. Jets carrying radiation are ``heavier''.

\begin{figure} 
\epsfxsize=12.5 true cm
\epsfig{file=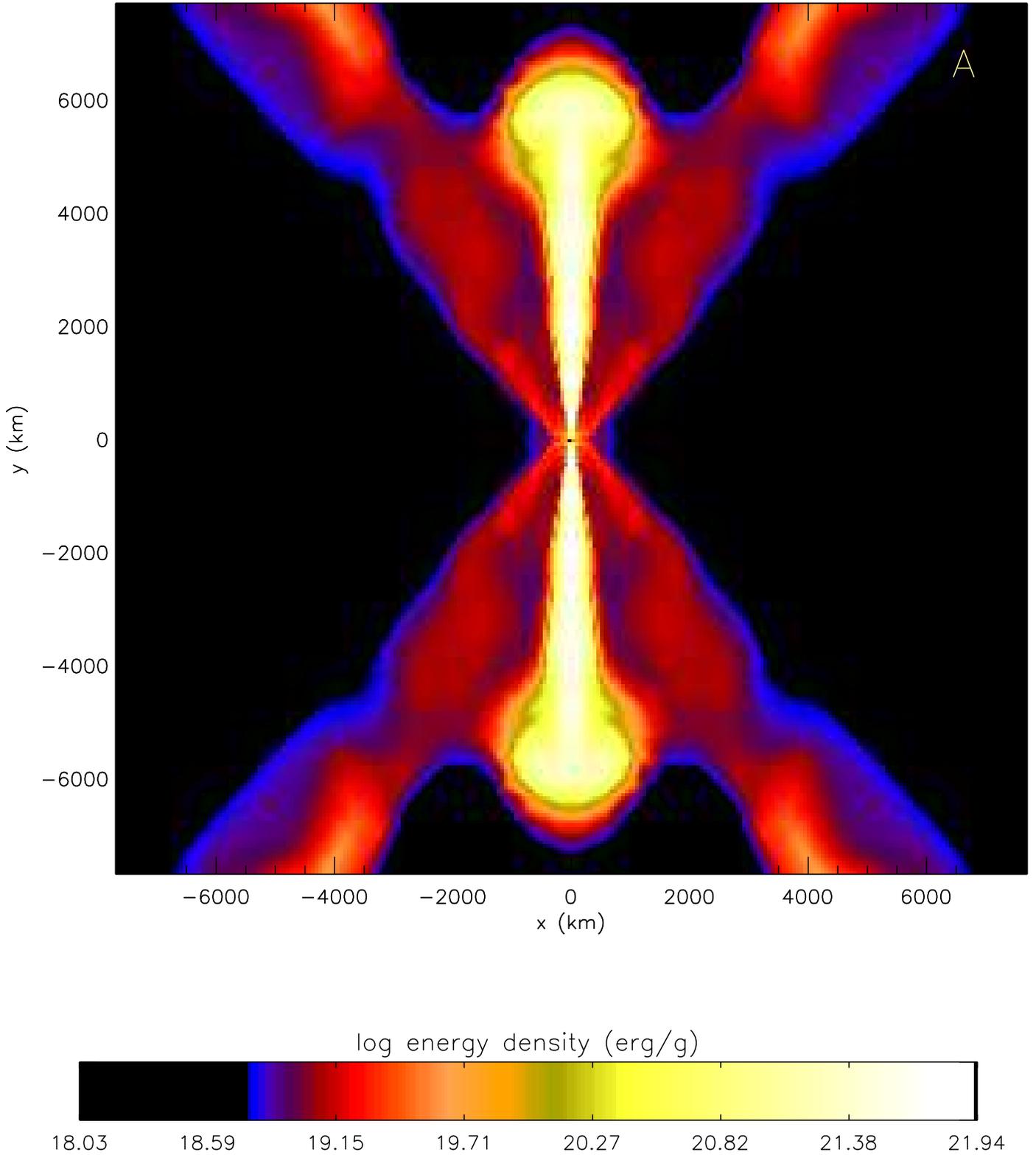}
\caption{\label{latejet}
Energy density in the jet and surrounding area 0.824 s after its
initiation. The jet has now moved 7,000 km. While its velocity cannot
be accurately determined in the non-relativistic code used for this
calculation, the energy density is typical of highly relativistic
matter with $\Gamma$ over 10. The jet remains highly focused with an
opening angle (half width) of about 10 degrees.  The red regions at
polar angles of 35 degrees are the plumes formed earlier by
dissipation in the disk ($\S$4.1.5 and Fig.\ref{plumes}). }
\end{figure}

Still we followed our non-relativistic jet awhile longer. 0.83 s after
energy deposition began (350,000 time steps, at t = 8.42 s) the shock
had moved to 7000 km. Fig. \ref{pjet} shows the conditions at this
point.  The internal energy density is roughly constant at all but
very small radii and still several times 10$^{21}$ erg g$^{-1}$
indicating mildly relativistic matter. A plot of entropy per baryon
(not shown) would look very similar to that for the internal energy
per gram, but with a value behind the shock of about 10$^4$. The speed
of the shock front has declined by this point to just over 10,000 km
s$^{-1}$, but for the reasons stated in the last paragraph we think
this is an underestimate of the real value.  Outflows of $\sim$ 50,000
km s$^{-1}$ have developed at intermediate angles between the polar
jet and the accretion disk.  The large inversion in density at the
shock is chiefly a consequence of lateral expansion behind the shock
and only partly due to the snow-plowing of material just ahead of the
shock. Outside of $1.5 \times 10^9$ cm, the density, temperature, and
pressure structures are all the same as when the jet was initiated.
The pressure in the jet is high and the jump across the shock
correspondingly large. The temperature and pressure can be estimated
approximately from the fact that the most of the energy deposited in
the simulation goes into the internal energy of the jet. Thus $aT^4$
times the volume of the jet is about 10$^{51}$ erg. The volume (of two
jets) is $2 \pi r^3 \theta^2$ where $\theta$ is the full opening angle
of the jet, about 0.4 radians. Thus $T \approx 4 \times 10^9$ K and $P
\approx 10^{24}$ dyne cm$^{-2}$. Fig. \ref{pjet} shows that these are
good approximate values for the temperature and pressure, but there
are significant gradients in both. Fig \ref{latejet} shows the
structure at this time.

By 8.42 s the calculation had become unrealistic with speeds behind
the jet head appreciably superluminal. It was stopped. The study needs
to be done with a relativistic hydrodynamics code. Such
calculations are already in progress (Aloy et al. 1999), but we can already
make some observations from our preliminary study.

First, most of the energy deposited in the bubble, up until the time
that it breaks out, goes into driving its expansion. Pressure and
density gradients are such that the bubble remains very elongated -
``focused''. When it breaks through the surface of the star, and we
estimate that it will in roughly 5 seconds, the evacuated channel will make
a collimated path for the unhindered escape of what is essentially a
pair fireball. This beam will be relativistic and highly focused.

The work that the bubble does in expanding, essentially PdV, goes into
displacing the overlying matter. This energy, which is quickly shared
by a lot of matter, will power an general (albeit asymmetrical)
explosion of the star. The total value can be easily calculated. It is
just the rate of energy deposition at the base times the time it takes
the jet to break out -- roughly 3 $\times 10^{51}$ erg. The work done
against gravity is a small subtraction.

As the leading edge of the bubble/jet breaks through the surface, the
escaping matter will be further accelerated by shock steepening in the
density gradient. This shock break out (e.g., McKee \& Colgate 1973;
Matzner \& McKee 1998) marks the first possible detection of the
explosion even though the core collapse occurred $\sim$5 s earlier.
Compared to the GRB, this prompt emission is probably faint, but some
hard emission - below the pair creation threshold - may occur as the
relativistic matter makes first contact with the surrounding
circumstellar medium. These hard x-rays travel faster even than a
$\Gamma$ = 100 jet and, by a radius of $3 \times 10^{15}$ cm, lead it
by $\sim$10 s. This may be the origin of hard x-ray precursors
sometimes seen in GRBs.

\section {GAMMA-RAY BURSTS}

According to current views, the principal GRB is made either as the jet
encounters roughly $\Gamma^{-1}$ of its rest mass in circumstellar or
interstellar matter (e.g., Rees \& M\'esz\'aros 1992) or by internal shocks
in the jet (e.g., Rees \& M\'esz\'aros 1994). If our jet has total energy
equivalent to an isotropic energy of 10$^{53}$ erg and $\Gamma$ of 100, it
will lose its energy after encountering $5 \times 10^{-6}$ M\sun \ (actually
this value times the beaming fraction).  If the star before the explosion was
losing 10$^{-5}$ M\sun \ y$^{-1}$ at 1000 km s$^{-1}$, the burst will be
produced at a radius of about 10$^{15}$ cm. Its duration will then be $R/(2
\Gamma^2 c) \sim 1$ s for $\Gamma \sim$ 100.

However, our jet is produced over a longer time than 1 s, so its
duration will not be governed solely by light propagation effects, but
by the time the engine operates after the polar regions have
cleared, about 10 - 20 s. Moreover, ours is an unsteady jet. Thus the GRB
will have time structure given not only by the circumstellar
interaction, but also by any observable residuals from the unsteady
flow.

\subsection {Time structure}

\subsubsection {Internal shocks}

Rees \& M\'esz\'aros (1994) describe the production of a GRB by unsteady
outflow. For two relativistic factors, $\Gamma_1$ and $\Gamma_2$
emitted in the jet $\Delta t$ apart, an internal shock will form at a
distance $\Gamma_1 \Gamma_2 c \Delta t$, releasing a significant
fraction of the energy in the jet. For the very rapid time variation
in Fig. \ref{mdot}, especially the 50 ms power peak, and for $\Gamma \ltaprx
100$, shocks will form at $\ltaprx$10$^{13}$ cm. This is too small a
radius for high energy gamma-rays to escape without producing an
optically thick fireball. Panaitescu et al. (1997) give an approximate
``thinning radius'', $r_t = 1.9 \times 10^{13} \, E_{51}^{1/2}
(100/\Gamma)^{1/2}$ cm, where the fireball becomes optically thin to
Thomson scattering. The energy one should use in this expression is
the equivalent isotropic energy, $E_{51} \sim 100$ for our
models. Thus the jet becomes optically thin at about $2 \times
10^{14}$ cm. Time structure shorter than about 1 s will be smoothed by
internal shocks happening internal to the $\gamma$-ray photosphere.

Interestingly the radius where one becomes optically thin and the
radius where the jet encounters $1/\Gamma$ of its rest mass are
comparable. So depending upon the actual mass loss history of the
presupernova star, one may get a combination of emission from internal
shocks and circumstellar interaction.

\subsubsection {A precessing jet?}

Woosley (1995) and Hartmann \& Woosley (1995) suggested, and
Portegies-Zwart, Lee, \& Lee (1998) have recently explored in some
detail, the possibility that some of the time structure observed in
GRB light curves may be due to precession of the black hole induced by
imperfect alignment of the black hole equator and the accretion
disk. The gravitomagnetic precession rate of the black hole is (Hartle
et al. 1986)
$$\Omega_{GM} \ \sim \  0.1 \left({{M_{disk}} \over {M_{\scriptscriptstyle
\odot}}}\right) \ \left({{M_{bh}} \over {5~M_{\scriptscriptstyle
\odot}}}\right)^{1/2} \ \left({{10^8 {\rm cm }} \over {\rm
r}}\right)^{2.5} \ \ {\rm s^{-1}},$$ 
which for a black hole mass of 3 M\sun, disk mass $\sim$1 M\sun \
(possible only for low viscosity), and disk radius 200 km gives a 
period of about a second.

Larwood (1998) gives a different expression for (non-relativistic)
forced precession
$$P_p \ = \ {{7} \over {3}} P_{\rm orb} {{(1 + \mu)^{1/2}} \over {\mu
\, {\rm Cos}\nu}} \left({{q} \over {r_{\rm disk}}}\right)^{3/2}$$
where $\mu$ = M$_{\rm disk}$/M$_{\rm bh}$ and $\nu$ is the angle
between the disk and axis of the rotating hole. For disk masses 1/3
that of the hole (certainly an upper bound appropriate to low
viscosity), radii of 200 km, and orbital periods $\sim$20 ms this
gives precession periods of about 0.2 s {\sl provided the angle
between the disk and black hole equator is significant}. Since the
black hole accretes most of its angular momentum from the disk we do
not expect this angle to remain large, even if the black hole was for
some reason born rotating obliquely (this seems unlikely). This
precession of the black hole would not necessarily be strictly
periodic since both the distance, b, and the mass of the disk are time
variable.

Additional structure in the time history and spectral hardness of the
burst would result from propagation effects. The highest $\Gamma$
material would be seen by a distant observer first even though, in a
symmetric pulse, lower $\Gamma$ crossed their line of sight earlier.
This would give a time asymmetry to a pulse originating from a beam
that was symmetric in $\Gamma$ about its central axis. 

Any effect that caused the jet to not be coalligned with the
rotational axis of the star would result in much greater baryon
loading and might quench the burst (and enhance the accompanying
supernova).  

\subsubsection{A traveling hole?}

The mechanism whereby pulsars receive a large ``kick'', typically
several hundred km s$^{-1}$, during, or shortly after a supernova
explosion remains uncertain. Prior to its collapse into a black hole
the central object in the collapsar briefly exists as a
protoneutron star - perhaps endowed with high rotation and a strong
magnetic field. Appreciable neutrino emission may occur prior to
collapse inside the event horizon. If for some reason that emission is
asymmetric, the black hole may acquire a kick. The magnitude is
presently impossible to estimate, but were it to be, e.g., 100 km
s$^{-1}$, the black hole would travel thousands of km during the
course of its accretion - a fraction of the radius of the stellar
core. The geometry of accretion and especially the focusing of the jet
would be affected, probably adversely since the accretion energy would be
shared by more mass. This gives yet another possibility for
diversity in GRB properties.

\subsection {Supernova 1998bw}

SN 1998bw was an unusual supernova in many ways (Galama et al. 1998). Models
that explain the observations (Woosley, Eastman, \& Schmidt 1998; Iwamoto et
al. 1998) require a very large kinetic energy, $\gtaprx 2 \times 10^{52}$ erg
if the explosion was isotropic, perhaps less if it was not (H\"oflich,
Wheeler, \& Wang 1998). High velocities for heavy elements are required to
explain the spectrum, about 10$^{49}$ erg of mildly relativistic ejecta to
explain the radio (Kulkarni et al. 1998a; though see Waxman \& Loeb 1998),
and the ejection of $\ltaprx$0.5 M\sun \ of $^{56}$Ni to power the light
curve. This supernova was also accompanied by an unusual GRB (GRB 980425;
Galama et al. 1998) which had only 10$^{48}$erg of gamma-rays (times a
beaming factor probably much less than one), lasted about 20 s, and had very
little emission above 300 keV. This is about 5 orders of magnitude less
energy than GRB 971214 (again depending on beaming) and other GRBs for which
red-shifts have been determined. Because the burst was not unusually bright
and yet so nearby (38 Mpc), there may be many more bursts like this that have
gone undetected. They could dominate the GRB source distribution at
sufficiently low fluence and show up as an isotropic unbounded set.

Within the context of the collapsar model, SN 1998bw/GRB 980425 was
the collapse and partial explosion of a massive helium star much like
Model 14A, but in which, for reasons to be discussed, the component of
the relativistic jet directed along our line of sight was weak (Woosley
et al. 1998). It was a powerful explosion nevertheless, probably
of the same order of magnitude as the one that made GRB 971214, and certainly
asymmetric. Depositing 10$^{52}$ erg, by whatever means, deep inside an
object as deformed as Fig. \ref{dens14a.big} will naturally lead to an
asymmetric explosion with higher velocities in a smaller amount of matter
along the rotational axes. But insufficient energy (or insufficient time)
may have existed in SN 1998bw to make a 10$^{52}$ erg jet (1.5\% beaming) with
$\Gamma \gtaprx 100$.

We believe that supernovae like SN 1998bw are generic to all
GRBs, but that in other GRBs with optical counterparts the event
was so far away and the relativistic jet in our direction so powerful
that the supernova was obscured by the optical afterglow from
shock deceleration.

In the collapsar model, the supernova is powered by two
sources. First, and probably most important, is the energy deposited
by the jet ($\S$4.1.7; Fig. \ref{latejet}) as it initially penetrates
the star. This energy is roughly the mass still contained within the
beaming angle of the jet times the square of the velocity with which
it is displaced. For Model 14A this is about 0.1 M\sun \ (including
both poles) times 1\% to 10\% mc$^2 \approx$ a few $\times 10^{51}$
erg. As this displaced material moves away, supersonically, from the
rotational axes, explosive nucleosynthesis occurs in the deeper
regions, producing some $^{56}$Ni to power the supernova light
curve. An additional source of supernova energy and of $^{56}$Ni is
the wind driven by viscous interactions in the disk (e.g., Katz 1997,
represented here by disk viscosity, $\S$4.1.5, Figs. \ref{plumes} and
\ref{plumee}). Some of this mass ejection is at high velocity,
especially the closer one goes to the poles. There the velocity also
increases in the steepening density gradient near the surface of the
star and becomes mildly relativistic ($\Gamma \approx 3$). The
circumstellar interaction of this material made GRB 980425 (McKee \&
Colgate 1973; Woosley et al. 1998; Matzner \& McKee 1998).

After the jet breaks through the surface, if enough time and energy
remain, the relativistic $\Gamma$ of the outflow increases
dramatically as the flow becomes unconfined. This is the
stage in which a ``classical'' GRB can be produced, but probably
was not in SN 1998bw. Or if it was we were not well situated to see it.

\subsection {GRB-971214}

At the other end of the spectrum of GRB diversity we have GRB-971214
(Kulkarni et al. 1998b), roughly 10$^5$ times more energetic in gamma-rays
than GRB 980425, with a harder spectrum (though similar time scale), an
optical ``afterglow'' that did not resemble a supernova, and a much lower
event rate in the universe. Can the collapsar model explain both?

We believe that the collapsar produces strong, hard GRBs like GRB
971214 only in the most extreme cases of high accretion rate and long
duration - perhaps only for the most massive stars or those that have
just the right angular momentum distribution. The jet must finish the
evacuation of the rotational axis of the star that accretion only
began. Once that has occurred, and that may take a few seconds, 
we speculate that a very powerful jet with low mass loading will begin
to blow. Energy is not such a problem.  Our standard model gives about
10$^{52}$ erg (for optimistic neutrino physics) focused into 3\% of
the sky and a duration of $\sim$15 s. This matches the observed
properties of GRB 971214 pretty well. Presumably there was also a
supernova underlying GRB 971214, but it was too far away to see and
fainter than the afterglow produced along our line of sight by the
relativistic jet.

However, baring some selection bias in which only the most energetic spikes
of an underlying enduring burst are seen, it does not seem possible for the
collapsar model to produce short, hard bursts. The group of bursts with mean
duarion 0.3 s (Fishman \& Meegan 1995) needs another explanation. These
bursts have a lot less energy than the long, complex bursts modeled here.
They may be the consequence of merging neutron stars or black hole - neutron
star mergers (Ruffert \& Janka 1998).

\section {CONCLUSIONS}

We have followed the evolution of rotating massive stars in which the
collapse of the iron core leads to the prompt formation of a black
hole. In essence, we have attempted to answer the question ``If
supernovae are the observational consequence of neutron star
formation, what then is the consequence of (prompt) black hole
formation?'' (see also Bodenheimer \& Woosley 1983; Woosley 1993). We
have demonstrated that the answer is ``a gamma-ray burst'' and
perhaps, ``a hypernova'' (Paczy\'nski 1998). The model that
makes the observable phenomenon called a hypernova, is the collapsar.

To simplify matters and because it makes the production of a GRB
easier, we have followed the evolution of bare helium cores, but our
results also carry over to stars that have not lost their hydrogen
envelope. Interesting phenomena await exploration there - an enduring
x-ray source not of a binary nature? A Type II supernova powered by
black hole formation?

Using the 14 M\sun \ helium core of a 35 M\sun \ main sequence star as
a prototype, we have begun to explore what may be a large parameter
space of mass, angular momentum distribution, and disk physics. Our
preliminary results show a new kind of phenomenon, a very energetic
stellar explosion of up to $\sim$10$^{52}$ erg, powered by
hyper-accretion into a black hole. Favorable geometry for jet outflow
develops as a consequence of the stagnation of matter in an equatorial
disk while matter along the rotational axes (initially) falls into the
hole (see also Woosley 1993; Jaroszy\'nski 1996). Lower mass
progenitors and higher angular momenta give, in our simplest neutrino
powered explosions, weaker bursts. Helium cores over 14 M\sun \ and
angular momenta down to half that studied here would probably give
even more powerful explosions that lasted longer.

The collapsar develops a GRB in stages and it may be that the sequence
does not always make it to completion. Powerful explosions may occur
in which the GRB is weak or absent. The first stage is the formation
of the disk and partial evacuation of the polar regions. This takes
several seconds.  While polar accretion continues at a high rate, a
jet cannot develop. So long as the density in the polar regions
exceeds $\gtaprx10^6$ g cm$^{-3}$, the inward momentum ($\rho c^3$
times area) dominates any reasonable energy deposition (10$^{51}$ erg
s$^{-1}$). Any energy added is advected into the hole and lost. After
a total of about 5 s though the pole does clear sufficiently that a
reversal of flow becomes possible.

The outflows that develop then have several origins. If the disk
viscosity is high ($\alpha \sim 0.1$), dissipation in the disk can
power a very energetic ``wind'' ($\S$4.1.5) that is almost
supernova-like in terms of mass, energy, and $^{56}$Ni
content. Energy deposition by neutrino annihilation can power polar
outflows, relativistically expanding bubbles of radiation, pairs, and
baryons focused by density and pressure gradients into jets. Because
the black hole rotates very rapidly at this point ($a \approx 0.9$),
MHD processes may also contribute to jet formation.

\begin{figure} 
\vspace{-.5 in}
\epsfig{file=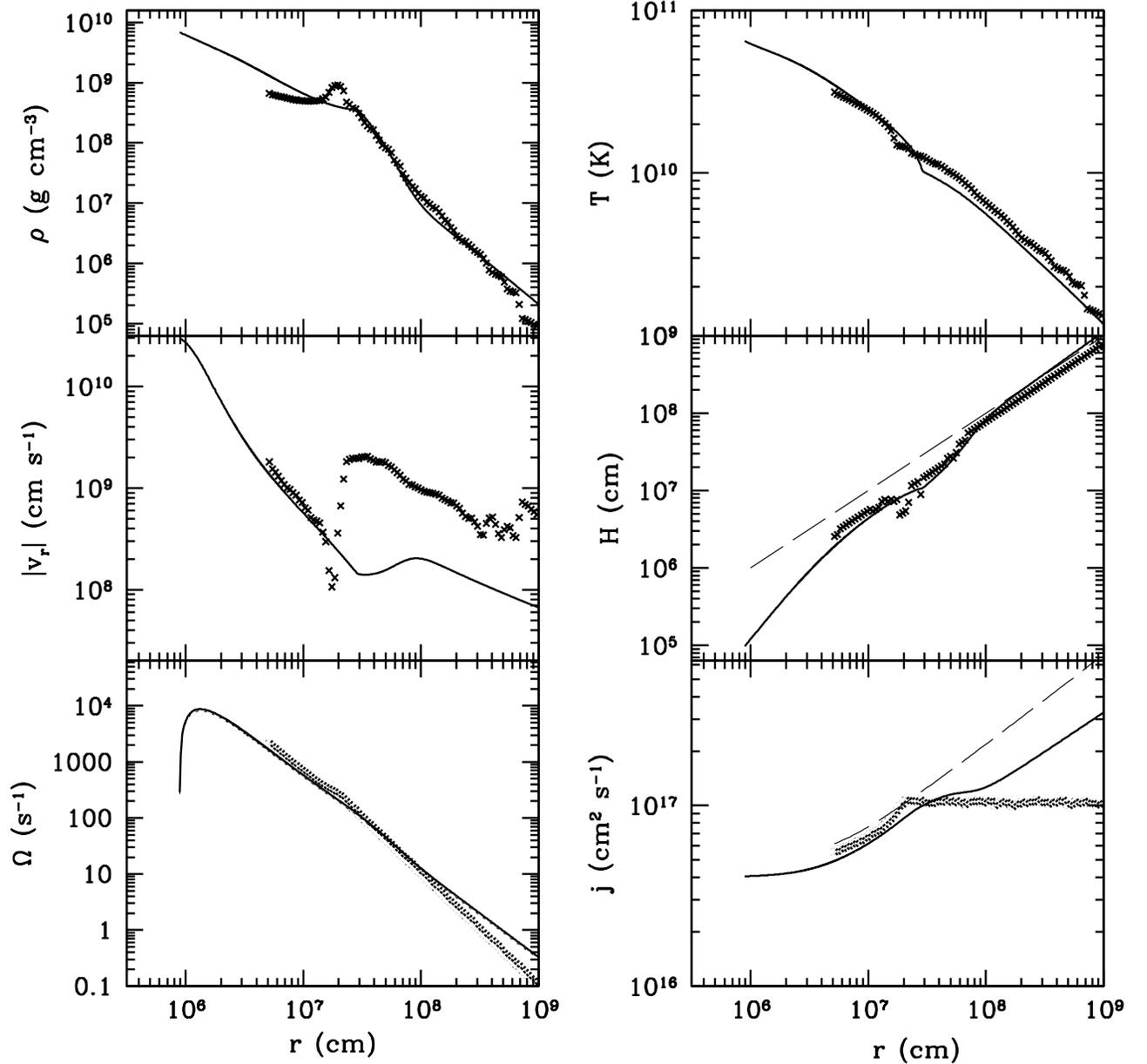,width=7 in}
\vspace{-.3 in}
\caption{\label{popham} The simulation (crosses; density, temperature,
radial velocity, disk scale height, angular frequency, and angular
momentum) at t = 7.39 s during a state of high mass
accretion rate (0.12 solar masses per second) compared with the
semianalytic solution of Popham, Woosley \& Fryer (1998; solid
lines). The equatorial values for the the 2D simulation are plotted
here. Interior to a transition region between 200 km and an accretion
shock at 250 km, the simulation matches the steady-state disc extremely
well. The temperature, to which the neutrino emission is
sensitive, is matched closely which gives us confidence in
using the total neutrino emission calculated by Popham et. al. for
their steady-state discs. The dashed line in the angular
momentum plot is the Keplerian value required to support the disc
entirely by centrifugal forces. The slight upturn at small radius is
an effect of general relativity.  The dashed line in the scale height
plot is the line H = r. The decrease below this line is due to
photodisintegration.}
\end{figure}

Because our numerical results are congruent with those of PWF in the
inner disk (Fig. \ref{popham}), we were able to use their analytic
models to estimate both the neutrino luminosity of our disk and the
efficiency for neutrino annihilation as a function of time
($\S$4.1.7). The result depends sensitively upon the accretion rate
and Kerr parameter $a$. For reasonable, but optimistic values, the
total neutrino energy emitted by the disk, during 20 s of accretion is
$3 \times 10^{53}$ erg (note the similarity in this energy to that
emitted in neutron star formation, but the luminosity here is about 10
times less. In neutrinos, these are {\sl not} the most powerful
explosions in the universe, though they are the brightest in photons).
The total energy deposited by neutrino annihilation was $1.4 \times
10^{52}$ erg. For less optimistic assumptions regarding the initial
Kerr parameter and the neutrino transport, the emitted energy was as
low as $1.4 \times 10^{53}$ erg and the deposited energy, $\ltaprx 1 \times
10^{51}$ erg. We emphasize that these numbers are for one particular
model, not chosen to be the optimal collapsar. Larger values are
possible, especially if the accretion rate is just a little higher.

We simulated this energy deposition ($\S$5) and followed its
consequences. Highly focused relativistic outflow - jets - developed.
After a fraction of a second, the energy to
mass ratio in these jets became very large, $\gtaprx 10^{22}$ erg
g$^{-1}$, corresponding to a large asymptotic relativistic $\Gamma$
factor.  The problem of ``baryonic contamination'' is circumvented
because the energy deposition blows a bubble of low density. Momentum
and energy from the annihilating neutrinos continues to be deposited
in this bubble even as the baryon fraction becomes small. This energy
is naturally directed outwards along the axis.  The head of the jet moves
much slower than the matter behind it that drives it. We followed this
jet through an appreciable fraction of the star's mass and radius and saw
that it maintained a collimation of about 10 degrees (half
angle). Certainly this part of our study needs to be redone using
relativistic hydrodynamics, but our results suggest that a sustained
jet is capable of breaking out of the star in $\sim$5 s, maintaining
collimation and relativistic speeds, with no great difficulty.

The jet expends a lot of energy though, perhaps several times
10$^{51}$ erg, clearing a channel through the
star. This energy goes into lateral expansion perpendicular to the jet.
Though we did not follow the explosion long
enough to see the complete disruption of the star, it is probable
that the accretion in the disk will be truncated at some point as the shock
warps around and ejects matter in the equatorial plane. This may not
happen for 10's of seconds though - the sonic crossing time. 
As the jet breaks through the surface, mildly relativistic matter is
for a range of polar angles down to $\sim$45 degrees. As this
matter runs into the precollapse mass loss of the star, a relatively
weak, soft GRB is created (Matzner \& McKee 1998; Woosley, Eastman, \&
Schmidt 1998). This may also be the origin of hard x-ray precursors
seen in some GRB's and of GRB 980425.

The principal GRB commences though only after the jet has broken out
of the star and continued long enough to evacuate a low density
channel for the relativistic plasma. This may take additional
time. Altogether it might be reasonable for the GRB producing jet to
commence 10 s after the black hole forms. Of course the GRB itself is
made far away from the star as the relativistic plasma runs into
material (or into itself) at several hundred AU ($\S$6). 

An intriguing new discovery is that the disk accretion rate while the
burst is being made is not steady. We calculate an accretion rate of
0.07 $\pm$ 0.03 M\sun \ s$^{-1}$ for about 15 s. The variations are
time resolved in the numerical study and have significant power on the
disk crossing time, $\sim$50 ms, but variations on all time scales up
to a few tenths of a second is seen. The disk instabilities that give
rise to these variations seem to be related to the location of a
region of nuclear photodisintegration just inside the accretion shock
that bounds the disk ($\S$4.1.4). 

\begin{figure}
\epsfxsize=12.5 true cm
\epsfig{file=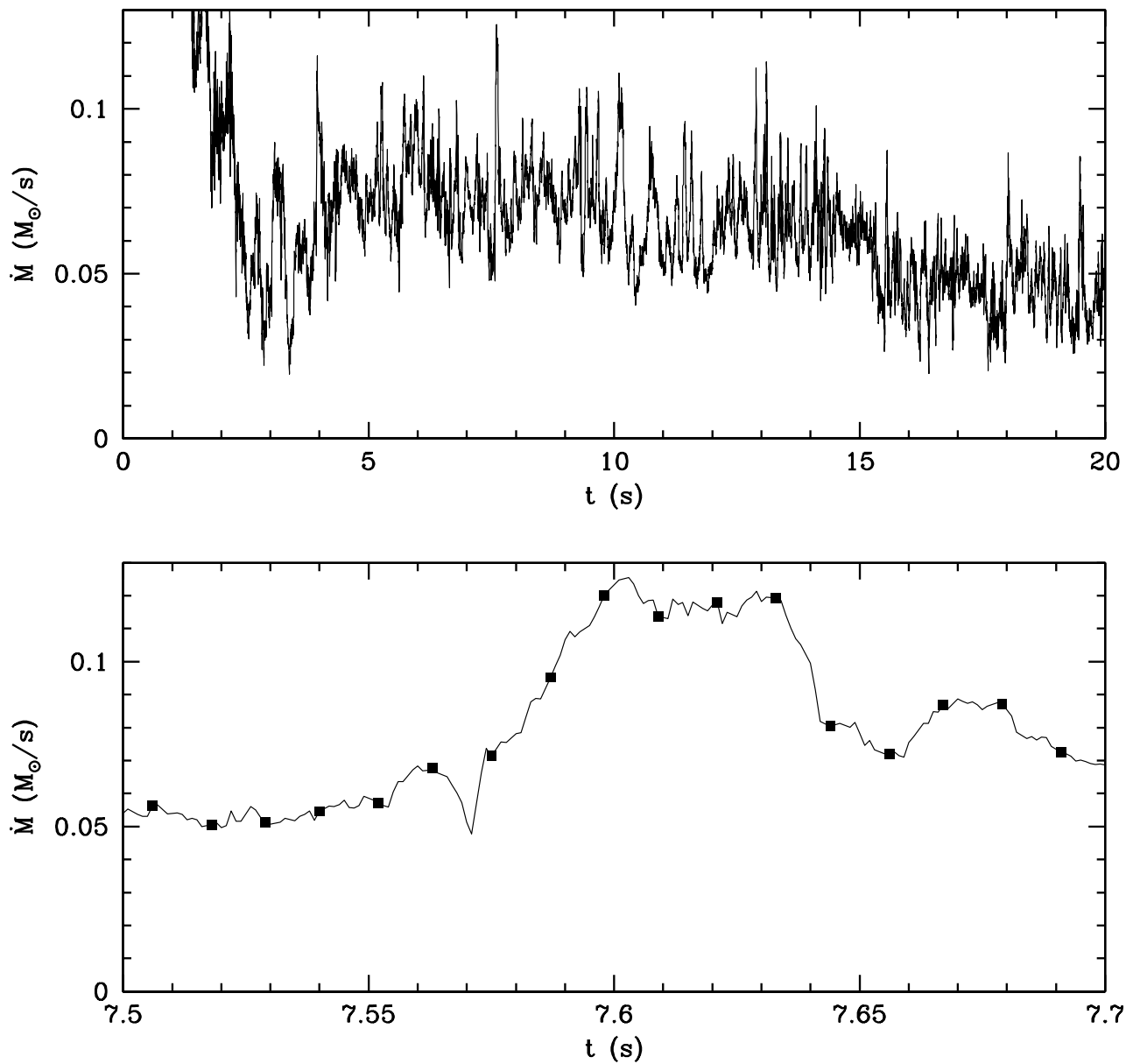,width=7 in}
\caption{\label{mdot}
Top panel: Mass accretion rate in Model 14A for the entire 20 seconds
the model was calculated within 45 degrees of the equator. After an
initial accretion transient during the first several seconds, the
accretion below 45 degrees is mediated through the accretion disc. An
average rate of .07 solar masses per second during the next twelve
seconds is modulated by fluctuations of more than 30\% on time scales
between tens and hundreds of milliseconds.  After 15 seconds the
average accretion rate dips below .05 solar masses per second. Bottom
panel: A mass accretion spike near 7.6 seconds representing a high
state of about 0.1 solar masses per second.  The black boxes indicate
times when full model dumps were written for analysis of the flow
variables. The time between 7.5 and 7.7 s is spanned by 17,500 model
calculations.}
\end{figure}

\begin{figure}
\epsfxsize=12.5 true cm
\epsfig{file=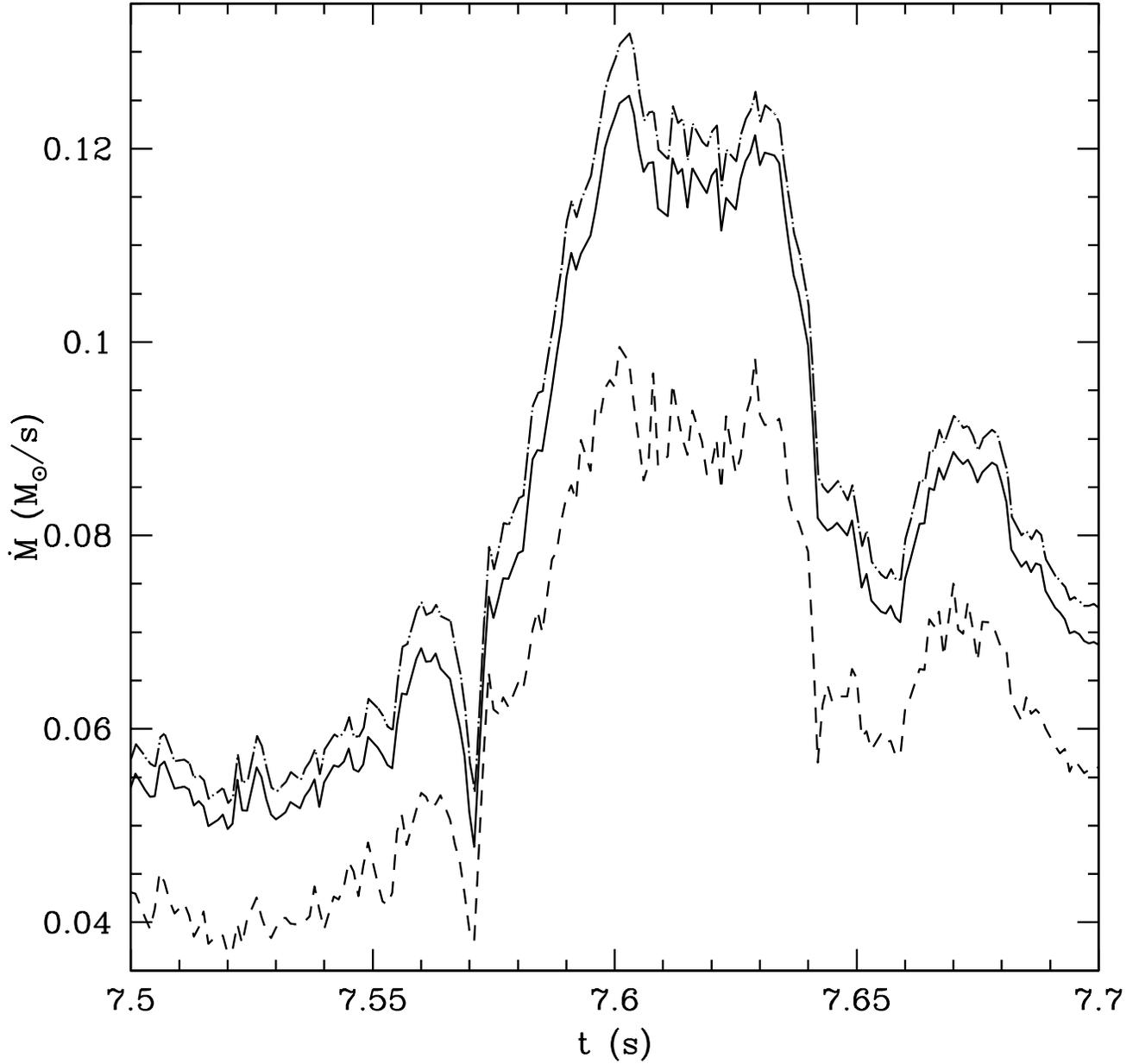, width=7 in}
\caption{\label{mdottheta}
Mass accretion rate as a function of angle during the episode
highlighted in Fig. \ref{mdot}. The integrated accretion rate is given
between the equator and 22.5 degrees (dashed line); 45 degrees (solid
line); and 90 degrees (dash-dotted line). Most of the accretion is through
the disk and not from the polar accretion column. }
\end{figure}

\begin{figure}
\epsfxsize=12.5 true cm
%\epsffile[0 0 612 612]{fourier.eps}
\epsfig{file=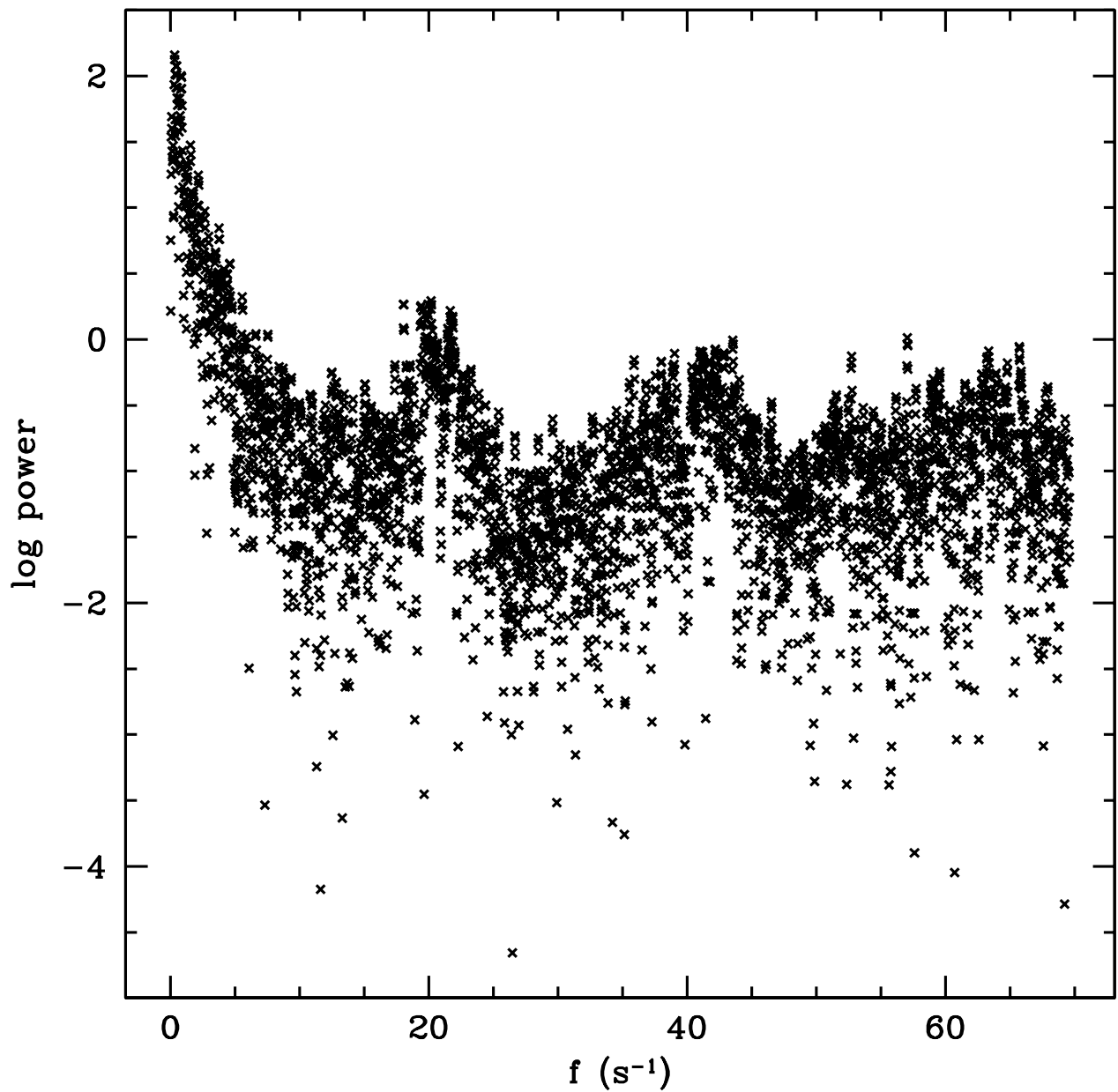,width=7 in}
\caption{\label{fourier} Fourier power spectrum of the accretion rate
shown in Fig. \ref{mdot}. Significant power peaks are seen at time
scales of 50 ms and 25 ms. The former is approximately the viscous
time of the disk and also approximately the orbit time at the outer
edge of the disk. The latter is its first overtone. The sampling of
models every 100 timesteps has a characteristic time scale of $\sim$ 1
ms (the Courant condition in the innermost disk set the timestep at
$\sim$ 10 $^-5$ s), so even time scales of 25 ms were well sampled in
the calculation. }
\end{figure}

\begin{figure} 
\epsfxsize=12.5 true cm
\epsfig{file=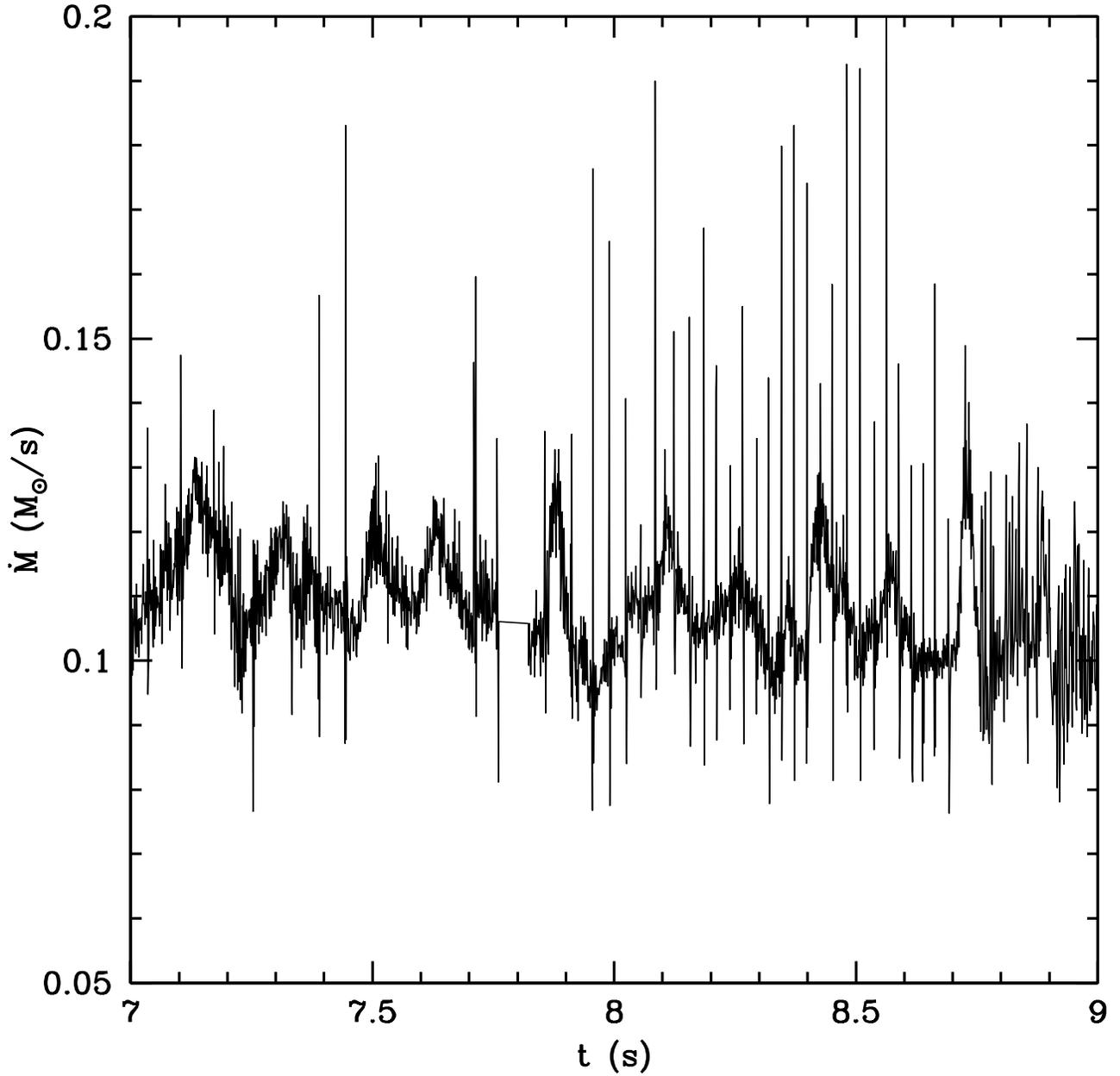,width=7 in}
\caption{\label{mdotloalp}
Despite its low viscosity, the accretion rate in Model 14B is about
the same as in 14A (Fig. \ref{mdot}). Mass flow through the disk is in
steady state with the rate at which it is being fed by stellar
collapse. Unlike Fig. \ref{fourier}, a Fourier power spectrum of this
rate shows no preferred frequencies. The narrow spikes are numerical
noise.}
\end{figure}

Because the accretion rate our model finds is coincidentally poised on
the knife edge of advective dominated and neutrino dominated disks
(PWF), the efficiency of neutrino deposition is highly time variable,
much more so than the small variations in accretion rate might lead
one to think. On time scales of 50 ms to 300 ms the jet essentially
turns on and off many times. Even these short time scales are long
compared to light and sound crossing times, so the jet responds almost
instantaneously. One thus expects the energy of the jet, i.e., its
$\Gamma$, to be highly time variable. Models in which the GRB is
produced by internal shocks in the jet are thus favored. Some of the
time structure in Figs. \ref{mdot} and \ref{edep} will be washed out
because of collisions in the jet interior to the gamma-ray
photosphere, but some may survive to produce the complex time history
on a scale of $\sim$0.1 - 1 s.

This leads us to a key difference between the collapsar model and
e.g., merging neutron stars and neutron stars plus black holes. The
merging compact objects release all their energy in a time of order 20
ms, i.e., short compared to the duration of most GRB's and result in a
thin shell of relativistic matter. Any time structure in the GRB
longer than about 10 ms thus reflects the circumstellar (or internal
shock) interaction and light travel delays. The collapsar, on the
other hand, is incapable of producing events shorter than about 10
s. It ejects matter not so much as a thin shell, but more like an
intermittent nozzle. Much of the time structure in the GRB light curve
may reflect conditions in the engine itself (i.e., the inner disk
inside 200 km). Shorter bursts can only result in the collapsar model
as a consequence of ``seeing the tip of the iceberg'' in what is
actually an enduring, albeit fainter underlying burst with complex
time structure. An exception might occur if the jet here precessed
($\S$6.1.2), but otherwise the existence of short hard bursts with
mean duration $\sim$ 0.3 s (Fishman \& Meegan 1995) may require
merging compact objects for their explanation (Janka \& Ruffert 1996;
Ruffert \& Janka 1998).

We have pointed out that the {\sl total} energy inferred for the
brightest known GRB to date GRB 971214; about $3 \times 10^{53}$ erg
in gamma-rays times beaming factor) and the faintest (GRB 980425;
$\sim 10^{48}$ erg) are really not that different when all reservoirs
of energy are taken into account. The kinetic energy of the supernova
(hypernova?) accompanying GRB 970425 was about 10$^{52}$ erg (Woosley,
Eastman, \& Schmidt 1998; Iwamoto et al. 1998; though see also
H\"oflich, Wheeler, \& Wang 1998). With beaming, the energy of GRB 971214 was
probably also $\sim$10$^{52}$ erg. The difference of course was how
much energy went into making gamma-rays beamed in our direction.

During its propagation through the star, the jet deposits enough
energy to explode, eventually, all the star that has not already
collapsed to the disk. Lacking a full special relativistic calculation
of the entire event, it is difficult to say exactly how much energy
this is, but rough estimates ($\S$5) give a few $\times 10^{51}$
erg. If the jet does not stay focused in a single direction, even more
energy may be deposited. Then there is the wind produced by the
viscous interaction in the disk ($\S$4.1.5), also $\sim$10$^{51}$ erg,
plus other sources of energy (MHD) not modeled here. So the star
explodes, and a total energy $\sim$10$^{52}$ erg may not be
unreasonable. If a powerful jet continues (and stays focused) for
another roughly 10 s after it has broken through the surface then a
powerful GRB like that of 971214 may result, provided we are looking
straight down the jet. Events like SN 1998bw may be more common
though, and may be seen even more frequently because of the larger
beaming angle. And finally GRB 980425 may have been a lot brighter in
gamma-rays had we viewed it more nearly pole on.

{\sl We predict that all GRB's produced by the collapsar model will
also make supernovae like SN 1998bw}. This is most likely those bursts
that last longer than a few seconds. For bright GRB's like 971214, the
optical afterglow from jet deceleration may obscure the fainter
supernova. Only a small fraction of supernovae make GRB's, but it may
be that most GRB's make supernovae.

This brings us to the event rate and spatial distribution of
collapsars. Clearly collapsars should be directly associated with star
forming regions. They are the deaths of the most massive stars and
should therefore be found only where such stars are being born. We
have suggested that there may be some metallicity dependence as
well. Lower metallicity makes the larger helium cores needed by the
collapsar model more likely and may also diminish the loss of angular
momentum. Still, even where they occur, we expect collapsars to be a
small fraction of supernovae (or any subclass thereof, e.g., Type
Ic). This is because the required mass is high and the requisite
angular momenta may not always be present. A value of a few per-cent
of the supernova rate, or 10$^{-4}$ y$^{-1}$ in a Galaxy like the
Milky Way seems reasonable. Even with beaming this would provide
enough GRB's to satisfy the observations of the bright events that
dominate the BATSE statistics. The number of fainter bursts like GRB
980425 would be much larger, but a) are only a small fraction of the
current BATSE data base and b) emit to a larger solid angle.  A more
detailed study of the event rate of this and other models is in
progress (Fryer, Hartmann, \& Woosley 1998).

An interesting implication of the collapsar model, if it is to explain
most GRB's, is that at least a fraction of iron core collapse events
in massive stars produce a neutron star that is initially rotating
almost at break up. It may be that this is the common case. If so
rotation would need to be included in models for ordinary supernovae,
not just those for GRB's. 

Finally, though relatively rare, collapsars may provide an important
nucleosynthetic component. We have not tracked with any care the
composition of the accretion disks or jets studied here, but the
entropy per baryon of the jet in Fig. \ref{pjet} is about
10$^4$. Smaller values will characterize the mass ejection at larger
angles. Our numerical resolution was inadequate to say just how much
disk material is mixed into the jet and we have also not followed the
evolution of the electron mole number, $Y_e$, in response to electron
capture in the disk. But if even 10$^{-5}$ M\sun \ of material with
$Y_e \ltaprx 0.4$ is ejected with entropies $\gtaprx$300, the
contribution to the $r$-process would be significant (Hoffman,
Woosley, \& Qian 1997).

\acknowledgments 

This research has been supported by by NASA (NAG5-2843 and MIT SC
A292701), and the NSF (AST-97-31569 and the International Program).  We
acknowledge many helpful conversations on the subjects of gamma-ray bursts
especially with Peter Bodenheimer, Chris Fryer, Dieter Hartmann, Alexander
Heger, Thomas Janka, Ewald M\"uller, Bob Popham, and Max Ruffert.  We
especially appreciate the calculation of Table 1 by Fryer and Popham. We
thank Steve Balbus, Xingming Chen, and Jim Stone for discussions of disk
viscosity and John Papaloiziou and Doug Lin for help in understanding disk
instabilities. We are particularly grateful to Bruce Fryxell who
provided an early version of the PROMETHEUS code used in these calculations.
A portion of this work was carried out at the Max Planck Institut f\"ur
Astrophysik in Garching and we gratefully acknowledge the support, good
company, and fine beer.

\newpage

\begin{table*}[t]
\begin{center}
\centerline {TABLE~1}
\vskip 8pt
\begin{tabular}{llllllll}
\hline\hline
\multicolumn{5}{c}{Conservative}&\multicolumn{2}{c}{Optimistic}\\
$\dot M$ & $a$ & L$_{\nu}$ & L$_{\nu \bar \nu}$ & efficiency & 
L$_{\nu}$ & L$_{\nu \bar \nu}$ & efficiency \\
M\sun \ s$^{-1}$ &  & 10$^{51}$ erg s$^{-1}$ & 10$^{51}$ erg s$^{-1}$
& \% & 10$^{51}$ erg s$^{-1}$ & 10$^{51}$ erg s$^{-1}$ & \% \\
\hline
0.05   & 0.50  &  1.2   & 0.00023 & 0.019    & 1.6  & 0.0012& 0.075 \\
0.05   & 0.75  &  2.2   & 0.0012 & 0.055   & 3.6  &0.016 & 0.44 \\
0.05   & 0.89  &  4.3   & 0.017  & 0.41    & 8.6  & 0.18  & 2.1  \\
0.05   & 0.95  &  7.6   & 0.061  & 0.81    & 18   & 1.3   & 7.4  \\
0.0631 & 0.95  &  23    & 1.9    &  8.2    & 35   & 3.7   & 10   \\
0.0794 & 0.95  &  35    & 1.9    &  5.3    & 39   & 2.1   & 5.3  \\
0.1    & 0.50  &  6.1   & 0.0083  & 0.14     & 7.8  & 0.027 & 0.34  \\
0.1    & 0.75  &  13    & 0.071  & 0.56    & 18   & 0.27  & 1.6  \\
0.1    & 0.89  &  33    & 1.2    &  3.6    & 36   & 1.2   & 3.5  \\
0.1    & 0.95  &  41    & 1.3    &  3.2    & 46   & 1.7   & 3.6  \\
\hline\hline
\end{tabular}
\end{center}
\end{table*}

\newpage

\clearpage

\clearpage

\begin{figure}
\vskip .5 in
\caption{\label{accshock.j} The velocity field in Model 14A at 0.751 s
showing early disc formation as centrifugal force begin to halt
accretion near the equator and a toroidal accretion shock forms at
about 550 km.  Meanwhile accretion along the polar axis proceeds
relatively uninhibited. The largest infall velocities just outside the
accretion shock in the equator are about 25,000 km s$^{-1}$
and at the pole 55,000 km s$^{-1}$. Material impacting the
disk at high latitude in channeled into the accretion column. Some
circulation in the disk is apparent. Because velocity arrows in this
and subsequent figures have their tails in the zone they represent,
the spherical inner boundary at 50 km is obscured.}
\end{figure}

\clearpage

\begin{figure}
\vskip .5 in
\caption{\label{accshock.abar} The velocity field at 0.751 s color
coded by composition. The color shows the logarithm mean atomic weight
of the nuclei in the accreting gas.  Here that weight ranges from 1 (nucleons)
to 24 (oxygen and silicon). The shock heats the infalling gas to
temperatures above 10 billion degrees causing heavy nuclei to
disintegrate.  The dark blue inner torus is composed of free neutrons 
and protons.}

\end{figure}

\clearpage

\clearpage

\clearpage

\clearpage

\clearpage

\clearpage

\begin{figure} 
\vspace{1 in}
\epsfxsize=17 true cm
%\epsffile[0 0 612 612] {mach.ps}
%\epsffile[0 0 612 612] {figure10.ps}
\vspace{-1.0 in}
\caption{\label{mach}
The radial Mach number, M$_r$, at 7.60 seconds in the standard run,
Model 14A, This figure highlights regions of supersonic flow and the
location of shocks, especially the one at the edge of the accretion
disk at 350 km.  The yellow arrows near the equator represent the gas
which feeds the disc (the tan toiroidal region interior to 350
km). The red arrows indicate outflow due to centrifugal bounce and
viscous heating.  }
\end{figure}

\clearpage

\clearpage

\clearpage

\clearpage

\clearpage

\clearpage

\clearpage

\begin{figure}
\vspace{1 in}
\epsfxsize=17 true cm
%\epsffile[0 0 612 612] {plumeform.ps}
%\epsffile[0 0 612 612] {figure17.ps}
\vspace{-1.0 in}
\caption{\label{plumeform} Origin of the viscosity driven ``wind"
shown in Fig. \ref{plumes}. Regions of high entropy (red; S $\approx
55$) are heated by viscous dissipation. Typical ratios of orbital
velocity to radial velocity are about 10 in the high entropy region
and the orbital speed is about 10$^{10}$ cm s$^{-1}$. Density in the
high entropy region is about $5 \times 10^7$ g cm$^{-3}$ and the
temperature, $1.5 \times 10^{10}$ K. Viscous dissipation in releasing
$ 5 \times 10^{29}$ erg cm$^{-3}$ s$^{-1}$.  All the material shown in
the figure, except that along the accretion column, is composed of
nucleons. When these reassemble the composition will be mostly
$^{56}$Ni and helium.}
\end{figure}

\clearpage

\clearpage

\clearpage

\begin{figure} 
\epsfxsize=12.5 true cm
\caption{\label{edep}
Energy deposition rate due to neutrino annihilation in the polar
regions for Model 14A assuming an initial Kerr parameter of 0.5 which
increased with time as in Fig. \ref{mbha}. The rates were calculated
by interpolating the models of PWF in mass accretion rate and Kerr
parameter. A constant black hole mass of 3 M\sun \ was assumed in this
interpolation. The top panel shows the time history of the energy
deposition; the middle panel, an expanded version of the energy
deposition from an accretion spike (Fig. \ref{mdot}). The lower panel shows
the total neutrino luminosity of the disc that gave the deposition in
the top panel. After about 15 s, the energy deposition declines due to
a decreasing mass accretion rate.}
\end{figure}

\clearpage

\clearpage

\clearpage

\clearpage

\begin{figure} 
\epsfxsize=12.5 true cm
\caption{\label{oxflow}
Velocity vectors show the onset of a mild explosion produced by a
combination of rotation and oxygen burning. The largest velocity
vector shown in the outflow is 17,000 km s$^{-1}$.}
\end{figure}

\clearpage

\clearpage

\begin{figure} 
\epsfxsize=12.5 true cm
\caption{\label{pjet}
Density, temperature, internal energy per gram, and pressure along the
polar axis (i.e., in the center of the jet) in Model 14A at a time of
8.43 s, 0.83 s after $5 \times 10^{50}$ erg s$^{-1}$ began to be
deposited just above and below the hole (see text). All quantities are
in cgs units. Pressure jumps about a factor of 100 in the jet. The
shock is very strong. Entropy in the shocked region is $\sim 10^4$.}
\end{figure}

\clearpage

\end{document}